\documentstyle[12pt,epsfig]{article}
\newlength{\dinwidth}
\newlength{\dinmargin}
\newcommand{\ba}{\begin{array}}
\newcommand{\ea}{\end{array}}
\newcommand{\beq}{\begin{equation}}
\newcommand{\eeq}{\end{equation}}
\newcommand{\bea}{\begin{eqnarray}}
\newcommand{\eea}{\end{eqnarray}}

\def\bce{\begin{center}}
\def\ece{\end{center}}

\def\nonu{\nonumber}

\def\al{\alpha}

\def\ep{\epsilon}

\def\La{\Lambda}

\def\eps6{{\displaystyle \mathop{\epsilon}^{6}}{}}

\def\nab6{{\displaystyle \mathop{\nabla}^{6}}{}}
\def\to{\rightarrow}

\newcommand{\bean}{\begin{eqnarray*}}
\newcommand{\eean}{\end{eqnarray*}}

\begin{document}

\thispagestyle{empty} \addtocounter{page}{-1}
\begin{flushright}
TIT-HEP-512 \\
{\tt hep-th/0312162}\\
\end{flushright}

\vspace*{1.3cm} 
\centerline{ \Large \bf More on ${\cal N}=1$ 
Matrix Model Curve for Arbitrary $N$ }
\vspace*{1.5cm}
\centerline{{\bf Changhyun Ahn }$^{\ast}$
\footnote{Address after Jan. 5, 2004:
School of Natural Sciences, Institute for Advanced Study, Olden Lane, 
Princeton, NJ 08540, USA  }
 and 
{\bf Yutaka Ookouchi }$^{\dagger}$}
\vspace*{1.0cm} \centerline{\it $^{\ast}$Department of Physics,
Kyungpook National University, Taegu 702-701, Korea}
\vspace*{0.2cm} \centerline{\it $^{\dagger}$Department of Physics,
 Tokyo Institute of
Technology, Tokyo 152-8511, Japan}

\vspace*{0.8cm} \centerline{\tt
ahn@knu.ac.kr,  \qquad 
ookouchi@th.phys.titech.ac.jp}
\vskip2cm

\centerline{\bf Abstract}
\vspace*{0.5cm}

Using both the matrix model prescription and the 
strong-coupling approach, we describe the intersections 
of $n=0$ and $n=1$ non-degenerated 
branches for quartic (polynomial of adjoint matter)
tree-level 
superpotential in ${\cal N}=1$ supersymmetric $SO(N)/USp(2N)$
gauge theories with massless flavors.
We also apply the method to the degenerated branch.
The general matrix model curve on the two cases 
we obtain is valid for arbitrary 
$N$ and extends the previous work from strong-coupling approach. 
For $SO(N)$ gauge theory  with 
equal massive flavors, we also obtain the matrix model curve on the
degenerated branch for arbitrary $N$. Finally we discuss on  
the intersections 
of $n=0$ and $n=1$ non-degenerated 
branches for equal massive flavors.

\baselineskip=18pt
\newpage
\renewcommand{\theequation}{\arabic{section}\mbox{.}\arabic{equation}}

\section{Introduction}
\setcounter{equation}{0}

\indent

The exact quantum effective
superpotential for the glueball field
was proposed by Dijkgraaf and Vafa \cite{dv1,dv2,dv3} using 
a zero-dimensional matrix model.   
Extremization of the effective glueball superpotential 
has led to the quantum vacua of the supersymmetric gauge theory.
For ${\cal N}=1$ supersymmetric $U(N)$ gauge theory with the adjoint 
matter $\Phi$, the gauge group $U(N)$ breaks into $\prod_{i=1}^n U(N_i)$
for some $n$ where $N=\sum_{i=1}^n N_i$. 
At low energies, the effective theory becomes 
${\cal N}=1$ gauge theory with gauge group $U(1)^n$.
The low energy dynamics of this gauge theory along the line of 
\cite{dv1,dv2,dv3}
have been studied in 
\cite{cdsw,csw,csw1,fer1,fer2,fer3,fo2}.

Although 
the ${\cal N}=2$ factorization problem in the strong-coupling approach
\cite{civ,cv}
has been solved 
for small values of $N$ explicitly \cite{csw}, the difficulty of the 
solving the general (arbitrary $N$) factorization problem 
occurs.
In \cite{Shih}, a matrix model curve 
for $U(N)$ gauge theory with cubic tree-level superpotential 
where the gauge symmetry 
breaks into $U(N) \to U(N_1)\times U(N_2)$ with $N=N_1+N_2$ 
was studied. 
One of the lessons 
is that the description of
minimization of glueball superpotential is useful to obtain 
a matrix model curve at $n=1$ and $n=2$ singularity \footnote{
For $U(N)$ gauge theory, the tree-level superpotential is written as 
$W(\Phi)=\sum_{k=0}^{n}\frac{g_k}{k+1}\mbox{Tr}\Phi^{k+1}$. 
Thus, the cubic case can be described by the value $n=2$. 
On the other hand, 
for $SO(N)/USp(2N)$ gauge theories, the tree-level superpotential 
can be written as 
$W(\Phi)=\sum_{k=1}^{n+1}\frac{g_{2k}}{2k}\mbox{Tr}\Phi^{2k}$. 
The quartic tree-level superpotential, which we will deal 
with mainly, is represented by the value $n=1$.} 
with {\it arbitrary}  $N$ for $U(N)$ gauge theory.
On the singularity, the two-branch cuts of the matrix model 
curve meet each other. In general, 
a matrix model curve can be represented 
as $y_m^2=W^{\prime}_2(x)^2+f_1(x)=
\prod_{i=1}^{4} \left(x-x_i\right)$ in terms of either a parameter 
of superpotential and fields or the roots of $y_m$. 
The period integrals on the Riemann surface can be 
written by the elliptic integrals. The glueball equations of motion 
allowed us to solve for the roots $x_i$ in terms of 
parameters $N_1, N_2$ and a parameter of superpotential.
It was not transparent to obtain the four different general roots $x_i$.
However at the singularity, 
since the root $x_2$ approaches the root $x_3$ (reducing of the roots of
$y_m$), 
by tuning the parameter of superpotential to special values depending 
on $N_1$ and $N_2$, it was possible to obtain the roots and parameter
of superpotential for general $(N_1,N_2)$.
This overcomes the difficulty in the strong-coupling approach 
(factorization problem) when $N$ is large because a matrix model 
curve is valid for {\it arbitrary} $N$. 

The extension \cite{csw} to 
the ${\cal N}=1$ supersymmetric gauge theories 
with the gauge groups $SO(N)/USp(2N)$
was obtained in \cite{ao}. The factorization problem gave an explicit
constructions for the matrix model curve for small values of $N$.
In \cite{ao2},  the description for 
$U(N)$ gauge theory  was generalized to the $SO(N)/USp(2N)$ gauge 
theories with quartic tree-level superpotential ($n=1$) by following the 
method of \cite{Shih}. 
In particular, a new result 
for degenerated branch
\footnote{
For $SO(N)/USp(2N)$ gauge theories, since a matrix model has ${\bf Z}_2$ 
identification the curve can be 
written in terms of  a function of $x^2$. 
One can 
observe that the position of the branch cut on the fixed point cannot move 
freely by changing a scale $\La$. 
If there are no branch cuts on the fixed point, called degenerated 
case, all the branch cuts move freely on 
the Riemann surface as illustrated in Fig. 4 in \cite{csw}. 
Therefore, the physics of these two 
cases are  different from each other. 
Thus, we have to study these two cases separately. 
However, for $U(N)$ gauge theory since there is no fixed point, we 
do not need to study them separately. This is one of the reasons why we
do not study $U(N)$ gauge theory with flavors in this paper.
It would be interesting to study $U(N)$ theory flavors in the future.} 
was obtained besides  
a simple generalization of the results in 
\cite{Shih} to these gauge groups. 
For the degenerated case, 
the number of roots of $y_m$ is less than the number of 
roots  on the non-degenerated 
case. The matrix model curve for the degenerated case 
\footnote{For the cases of 
$N_f = N - 3 $ for $SO(N)$ and $N_f= 2N + 1$ for 
$USp(2N)$ gauge theories, the extra piece proportional to $x^2$ should be 
added. See sections 3 and 5. }
can be written 
as\begin{eqnarray}
y_{m,d}^2=\left(\frac{W^{\prime}_3(x)}{x} \right)^2+
4F\equiv \left(x^2+x_0^2 \right)\left(x^2+x_1^2\right) 
\nonu
\end{eqnarray}
in terms of either a parameter 
of superpotential and field or the roots of $y_m$. 
Using a matrix model curve  for non-degenerated case, 
$y_m^2=\prod_{i=0}^{2} \left(x^2+x_i^2\right)$, 
the degenerated  curve can be related to 
$y_m^2=x^2 y_{m,d}^2$ implying that one of the roots 
in the non-degenerated case vanishes. This situation is quite 
similar to the one studied in \cite{Shih}. Therefore, 
the claim  in \cite{ao2} 
was that the glueball approach is suitable for computation of a general
matrix model curve on the degenerated case which depends on 
the field $F$, $N_0$ and $N_1$ for general 
breaking pattern $SO(N)\to SO(N_0)\times U(N_1)$ with $N=N_0+2N_1$. 
The $N$ can be any {\it generic} value.
We should not restrict our discussion to a 
special point on a matrix model curve but 
it covers the whole degenerated branch. Thus, as 
studied in \cite{ao} there exist some smooth interpolations 
between the vacua with different breaking patterns like as Coulomb 
branch in \cite{csw}. 

It is natural and interesting 
to describe the glueball approach by adding the massive or 
massless flavors into the 
pure $SO(N)/USp(2N)$ gauge theories. 
Recall that the Seiberg-Witten (SW) curves \cite{bl,ds,as} 
for $SO(N)/USp(2N)$ gauge theories with flavors
are characterized by 
\bea
y^2_{SO(N)} & = & P_{2[N/2]}^2(x) -4 \La^{2(N-2-N_f)} x^{2(1+\ep)} 
\prod_{f=1}^{N_f} \left(x^2-m_f^2 \right), 
\nonu \\
x^2 y^2_{USp(2N)} & = & \left( x^2 P_{2N}(x) 
+2 \La^{2N+2-N_f} \prod_{f=1}^{N_f}
m_f \right)^2 -4 (-1)^{N_f} \La^{2(2N+2-N_f)} \prod_{f=1}^{N_f}
\left( x^2 -m_f^2 \right)
\nonu 
\eea
where $\ep=0$ for $N$ odd and $\ep=1$ for $N$ even. 
When the number of flavors $N_f$ becomes zero in the power of $\La$
and the expressions containing the product for the mass $m_f$ become 1,
then the above formula  
reduces to the SW curve of pure gauge theory.
Compared with the pure case, the factorization problem in the 
strong-coupling approach with flavors becomes more complicated due to the
presence of flavor-dependent part. In the strong-coupling approach the 
${\cal N}=1$ matrix model curve is described by the single root part in the 
SW curve.
One of our aims is to observe the singularity on the matrix model curve 
(i.e.,double roots part) for {\it arbitrary} 
$N$ in our gauge theories \cite{afo1,afo2} from the point of
view of glueball approach.    
We expect to have a general matrix model curve for massless flavors 
without any difficulty 
because the analysis from the strong-coupling approach implies that
the SW curve looks like the curve for pure gauge theories  with different
power behavior of $x$ above.
However, for massive case (even equal mass for flavors), 
since we have an extra mass parameter in our factorization problem,
most of the matrix model curves for given $N$ and $N_f$ (the number
$r$ in the $r$-th vacua) in our gauge theories 
do not have an extra double root and therefore, they do not exist at 
the singularity we are interested in.
Instead, when they have an extra overall factor $x^2$ in the 
matrix model curve (i.e., degenerated case), there is a chance to 
obtain the general matrix model curve because the number of roots  is
reduced and this fact will make easier to compute the equation of motion for
glueball field.  

In this paper, we deal with two topics: 

$\bullet$ One is $n=0$ and $n=1$
singularity on the non-degenerated case. This is a simple 
generalization of the results in \cite{Shih} to the gauge theories
with flavors in the vector representation for $SO(N)$ gauge theory or 
fundamental representation for $USp(2N)$ gauge theory. 

$\bullet$ The other is a study for a general matrix model curve 
for the gauge theories on the whole degenerated branch. 

The effective superpotential for the gauge theories with some 
flavors was already discussed in 
\cite{csw1,NSW,Ookouchi,ow,ahn}. 
The contributions of the flavors 
to the effective superpotential, ${\cal F}_{flavor}$, can be expressed 
by the integral of matrix model curve (\ref{Fflavor}). In 
\cite{feng,afo1,afo2}, the mass of flavors possesses the same 
value and there exists one constraint, $W_3^{\prime}(\pm m_f)=0$. 
For the $SO(N)/USp(2N)$ cases, since the superpotential 
$W_4(x)$ is an even polynomial in $x$, 
the relation $W_3^{\prime}(x)=0$ 
always has a zero $x=0$ as a root. In other words, 
for massless case, the above constraint is automatically satisfied. 
Therefore, the massless case and the massive case are quite different 
from each other (One cannot obtain the massless case 
as we take the zero limit of 
the flavor mass in the massive case). 
As discussed in \cite{afo1,afo2}, the massless 
flavors are charged under the factor $SO(N_0)$, while 
the massive flavors are charged under the factor $U(N_1)$ in the 
breaking pattern $SO(N) \to SO(N_0) \times U(N_1)$. 
\footnote{Then, 
we use $\widehat{SO(N_0)}$ or $\widehat{U(N_1)}$ to represent the flavors 
which are charged under the corresponding gauge groups. Similar arguments 
hold for $USp(2N)$ gauge theory.}

The organization of this paper is as follows.
In section 2, 
we study the matrix model curve at $n=0$ and $n=1$ singularity for 
$SO(N)$ gauge theory with massless flavors. This is 
the simplest example, because the contributions coming from 
the flavors can be written in terms of 
the dual period integral $\Pi_0$. 
Therefore, using previous results obtained in \cite{ao2} for the 
pure gauge theory, we obtain a general matrix model curve (\ref{curve}) 
at the
singularity for the massless flavor case by minimizing the 
effective superpotential with respect to the glueball field. 
To demonstrate on the 
validity of general matrix model curve, we deal with some explicit 
examples for $SO(N)$ gauge 
theories where $N=4,5,6,7,8$ or 9 with $N_f(\le N-3)$ 
flavors. 
We have checked the precise values $N_0$ and $N_1$ explicitly 
by performing the period of $T(x)$ (characterized by
a characteristic function) over the compact cycles, for some
particular examples.
The addition map is used.

In section 3, we study a general matrix model curve on 
degenerated branch for $SO(N)$ gauge theory with massless flavors. 
Contrary to the
results in section 2, we obtain the matrix model curve characterized by
a single field. There are two classes of 
matrix model curves, (\ref{formula}) and (\ref{formu}), 
depending on  whether the quantity $(N_0-N_f-2)$ is equal 
to zero or not. After we obtain the general matrix model curve 
for arbitrary $N$ for $SO(N)$ gauge theory, 
we explicitly check the equivalence between 
two approaches (glueball approach and the factorization problem of 
SW curve). The subtlety for the matrix model curve is discussed when the 
number of flavor $N_f$ reaches 
the maximum value $(N-3)$ in the 
asymptotic region of the theory.
We predict 
the matrix model curve  for the  degenerated $SO(N)$ 
gauge theories with $N_f (\le N-3)$ massless flavors where $N \ge 7$
and $SO(N) \to \widehat{SO(N_0)} \times U(N_1), N_0 \neq 0$.

In sections 4 and 5, we extend the studies given in sections 2 and 3 
for $SO(N)$ gauge theories to the $USp(2N)$ gauge theory with massless 
flavors. 
The procedures are similar to the $SO(N)$ gauge theory. However, 
there exists one 
difference: the matrix model curves for the two gauge theories can be 
represented as $y_m^2=x^2\left(x^2 \pm m \right)^2+f_2(x)$ 
where the plus sign 
corresponds to  the $SO(N)$ gauge theory and the 
minus sign corresponds to  the 
$USp(2N)$ gauge theory. Therefore, by replacing 
$(N-2) \to (2N+2)$ and $x_i^2 \to -x_i^2$, 
we obtain the general matrix
model curves (\ref{curve1}), (\ref{curve3}), and (\ref{curve4})
for $USp(2N)$ theory from $SO(N)$ theory.
When the number of flavors $N_f$ reaches its maximum value $(2N+1)$ in a 
theory,
the analysis for the curve in degenerated case needs to have an extra term due 
to the flavor dependent term.  
We predict 
the matrix model curve for the  degenerated $USp(2N)$ 
gauge theories with $N_f (\le 2N+1)$ massless flavors where $N \ge 4$
and $USp(2N) \to \widehat{USp(2N_0)} \times U(N_1), N_0 \neq 0$.

In section 6, we study the degenerated 
$SO(N)$ gauge theories with massive flavors. 
Contrary to the massless cases, the constraint for 
the mass of flavors, 
$W_3^{\prime}(\pm m_f)=0$, is {\it not} automatically satisfied. 
However, this condition provides that the contribution arising from 
flavors can be written in terms of  a dual period $\Pi_1$. 
By using this fact and performing the equation of motion for 
glueball field, we obtain a general matrix curve for $SO(N)$ theory. 
Moreover, we discuss $r$-Higgs branch 
discussed in \cite{aps,Konishi,ckkm,afo1,afo2,csw1} in the context of
glueball approach. To obtain a general 
matrix model curve for this branch we follow the discussion given 
in \cite{csw1} about contour of integrals under the change of 
mass parameter and finally obtain the effective superpotential for the 
branch. Minimizing the effective superpotential we derive 
the matrix model curve for the theories (\ref{form}), (\ref{form1}), 
(\ref{tildeform}), and (\ref{tildeform1}). After the general 
discussions, we deal with some explicit examples, $SO(N)$ with 
$N=4,5,6$ and check the explicit agreement between 
two approaches. 
We predict 
the solutions for the  degenerated $SO(N)$ 
gauge theories with $N_f (\le N-3)$ massive flavors where $N \ge 7$
and $SO(N) \to SO(N_0) \times \widehat{U(N_1)}, N_0 \neq 0$.
\footnote{As one can see in \cite{afo2}, there is no 
degenerated branch for $USp(2N)$ theory. Therefore, we do not need to study
them here.}

In section 7, we give some comments on the $n=0$ and $n=1$ singularities 
for massive $SO(N)/USp(2N)$ cases. We do not obtain the general 
matrix model curves because, in general, the ${\cal F}_{flavor}$ cannot 
be written as a dual period $\Pi_1$. Therefore, the matrix
model curve cannot be represented by any simple formulation. However, 
when the condition, $c<0$ where $c$ is defined in (\ref{curvetilde}) 
or (\ref{curve6}), 
should be  
satisfied, the flavor-dependent part 
can be written  in terms of  $\Pi_1$. 
Thus, in this particular case, we 
obtain a matrix model curve for arbitrary $N$.

There exist many related works from the different matter 
representations for various gauge theories
\cite{intetal}-\cite{nsw1}.

\section{The $n=0$ and $n=1$ singularity: $SO(N)$ gauge theory with massless
flavors}
\setcounter{equation}{0}



\indent

Let us  consider $SO(N)$ gauge theory with quartic tree-level 
superpotential ($n=1$) and $N_f$ flavors. 
We are interested in 
 the intersections of the $n=0$ and $n=1$ branches.
At these special points, a vacuum with unbroken 
gauge group $\widehat{SO(N_0)} \times U(N_1)$ meets 
a vacuum with unbroken gauge group $SO(N)$ with 
$N=N_0+2N_1$.
Since we are considering an asymptotic free theory, 
we restrict the number of flavors $N_f$ to satisfy 
the condition $N_f< N-2$. As already clarified in 
\cite{afo1}, the relation between 
$W^{\prime}_3(x)=x\left(x^2+m \right)$ 
and $F_{6}(x)$ corresponding to an ${\cal N}=1 $ 
matrix model curve is characterized by   
\begin{eqnarray}
y_m^2 & = & F_6(x) =
W_3^{\prime}(x)^2+f_{2}(x), \ \qquad \qquad \qquad
N_f< N-3, \nonu \\
y_m^2 & = & F_6(x) =
W_3^{\prime}(x)^2+f_{2}(x)-4x^4\Lambda^{2}, \ \ \qquad
N_f= N-3 
\nonu
\end{eqnarray}
where $f_2(x) = f_2 x^2 + f_0$.
These relations are valid for any $SO(N)$ gauge 
theories where $N$ is even or odd. 
All masses of the flavors vanish. The $\La$ is an 
${\cal N}=2$ strong-coupling scale.
The fluctuating fields $f_2$ and $f_0$ relate to the glueball fields
$S_1$ and $S_0$ respectively.

The effective superpotential for the $SO(N)$ gauge theory with $N_f$ 
flavors is given by \cite{Ookouchi,NSW,ow,ahn} 
\bea
W_{\rm eff} = 2\pi i \left[ (N_0-2)\Pi_0+2N_1\Pi_1 \right]-2(N-2-N_f) S 
\log \left( \frac{\La}
{\La_0}\right) -{\cal F}_{flavor},
\nonu 
\eea
where the contributions from the flavors are given by the integral for
$y_m$ over $x$:
\bea
{\cal F}_{flavor}=\sum_{f=1}^{N_f}\int_{ \pm m_f}^{\La_0}y_m dx. 
\label{Fflavor}
\end{eqnarray}
The log-divergent term was renormalized by the bare Yang-Mills coupling 
and the cut-off $\La_0$ was replaced by the physical scale $\La$.
\footnote{For the tree-level 
superpotential of degree $2(n+1)$, the effective
superpotential for the $SO(N)$ gauge theory with $N_f$ flavors is
$W_{\rm eff} = 2\pi i \left[ (N_0-2)\Pi_0+\sum_{i=1}^n 2N_i\Pi_i 
\right]-2\left(N-2-N_f \right) \left( S_0 + \sum_{i=1}^n 2 S_i \right) 
\log \left( \frac{\La}
{\La_0}\right) -{\cal F}_{flavor}$. For quartic case ($n=1$), this will
lead to the above effective superpotential.}
Note that there exists a relation $S=S_0 +2 S_1$.
The periods $S_i$ and dual periods $\Pi_i$ of holomorphic 3-form
for the deformed geometry were written as the integrals over the 
$x$-plane \cite{ao2,fo,edel}.
The $m_f$ is the mass of the $f$-th flavor and we have chosen
the location of holomorphic 2-cycles at $x= \pm m_f$. Around these points,
D5-branes are wrapping on holomorphic 2-cycles, providing
a source for a single unit
of RR flux at $x= \pm m_f$. 
Thus the integral of 3-form RR flux around $x= \pm m_f$ is equal to 1.
The full effective superpotential 
can be generated by adding this contribution to the effective 
superpotential for pure gauge theory. 
 
First, let us consider the simple case in which all masses of flavors 
are zero in this section. Later, in sections 6 and 7, 
we will deal with the massive cases. 
Since the contribution from the flavors, 
${\cal F}_{flavor}$, becomes $ 2 \pi i N_f \Pi_0$, we can use the 
results for pure $SO(N)$ gauge theory studied in \cite{ao2} 
without much
efforts. 
We  only have to replace the quantity $(N_0-2)$ appeared in the 
matrix model curve in \cite{ao2} 
with $(N_0-N_f-2)$.   
Therefore, by combining the 
flavor-dependent part with the dual period $\Pi_0$ term,  
the effective glueball superpotential for massless flavors 
is
\begin{eqnarray}
W_{\rm eff}&=&2\pi i \left[ (N_0-2-N_f)\Pi_0+
2N_1\Pi_1 \right]-2(N-2-N_f) S \log \left( \frac{\La}
{\La_0} \right).
\nonu 
\end{eqnarray}
The periods 
of $T(x)$ defined as the periods of the 1-form at the $n=0$ and $n=1$
singularity
are given similarly and  together 
with different power behavior of $x$ and $\La$,
due to the presence of the flavors, 
the $T(x)$ also takes the form as a complete differential. 
With the above replacement in mind, we immediately obtain the 
matrix model curve near the singularity on the non-degenerated case from 
the glueball equations of motion: 
\begin{eqnarray}
y_m^2&=&x^2 \left(x^2+m \right)^2+
\langle f_2 \rangle x^2 +\langle f_0 \rangle =
\left[ x^2 + 2 \eta \La^2 \left(1+c \right) \right]^2 \left( x^2 +
4 \eta \La^2 \right), \nonu
\end{eqnarray}
where the fields, parameter of a superpotential, 
and the glueball field are given by 
\begin{eqnarray}
\langle f_2 \rangle &=& 4\eta^2 \La^4 \left(1+2 c 
\right),\qquad 
\langle f_0 \rangle=16\eta^3 \La^6 \left(1+c 
\right)^2, \nonu \\
m &=& 2\eta \La^2 \left(2+ c \right), \qquad  
\langle S \rangle=-\eta^2\La^4 \left(1+2 c 
\right), \qquad c \equiv \cos \left( \frac{2\pi N_1}{N-N_f-2} \right).   
\label{curve}
\end{eqnarray}
Here $\eta$ is the $(N-N_f-2)$-th root of unity for $(N-N_f-2)$ even and 
the $(N-N_f-2)$-th root of minus unity for $(N-N_f-2)$ odd. That is,
\begin{eqnarray}
\eta^{N-N_f-2} & = & 1,\qquad \ \  
\mbox{for}\ \ (N-N_f-2)\ \  \mbox{even},\nonu \\ 
\eta^{N-N_f-2} & = & -1,\qquad  \mbox{for}\ \ (N-N_f-2)\ \ \mbox{odd}. 
\nonu
\end{eqnarray}
This matrix model curve with massless flavors is exactly the same as pure case 
with the replacement $(N-2) \to (N-N_f-2)$. 
\footnote{One expects that the matrix model curve for $U(N)$ gauge theory 
with massless flavors ($U(N) \to U(N_1) \times U(N_2)$ where $N=N_1+N_2$) 
at $n=1$ and $n=2$ singularity takes the form 
$y_m^2 = \left[x + \eta \La (2+c)\right]\left[x-\eta 
\La (2-c)\right]\left( x-\eta \La c \right)^2$ where $c \equiv \cos 
\left( \frac{\pi N_2}{N-N_f} \right) $ and $\eta^{N}=1$. }
Note that the above solutions have the special combination $(N-N_f-2)$ 
which reflects the addition map discussed in \cite{afo1}, relating the 
Chebyshev branches (or the Special branches) of two different gauge groups. 
Namely, we have seen
that the Chebyshev branch of $SO(N)$ with $N_f$ massless flavors can be 
reduced to the Chebyshev branch of $SO(N^{\prime})$ with 
$N_f^{\prime}$ massless flavors when there exists a relation:
$N-N_f-2=N^{\prime}-N_f^{\prime}-2$.  

Let us demonstrate these general features (\ref{curve}) 
explicitly by comparing 
them with the results from strong-coupling approach 
obtained in \cite{afo1} already. Let us consider $SO(N)$ 
gauge theories with $N_f (\le N-3)$ flavors where $N=4,5,6,7,8$ or $9$.

$\bullet$ $SO(4)$ with $N_f=1$

In \cite{afo1}, the factorization problem 
resulted in the matrix model curve $\widehat{y_m}^2$
which contains an overall factor $x^2$. 
\footnote{We denote the curve from 
strong-coupling approach (factorization problem) by 
$\widehat{y_m}^2$ and the curve  from
glueball approach by $y_m^2$.}
Therefore, there exists no solution for non-degenerated case.
We will come to this  for degenerated case later.

$\bullet$ $SO(5)$ with $N_f=1$

In \cite{afo1}, the matrix model curve $\widehat{y_m}^2$
contains an overall factor $x^2$.
Therefore, there exists no solution for non-degenerated case.
We will come to this for degenerated case later.

$\bullet$ $SO(5)$ with $N_f=2$

The SW curve is the same as the curve of $SO(4)$ with $N_f=1$
which reflects the addition map.
Then the factorization problem will lead to 
the one in $SO(4)$ with $N_f=1$ theory.

$\bullet$ $SO(6)$ with $N_f=1$

This example is the first nontrivial case.
By putting the values $(N,N_1,N_f)=(6,1,1)$ into the curve (\ref{curve}), 
we predict the matrix model curve for this case,
\begin{eqnarray}
y_m^2=x^2\left(x^2+3\eta \La^2 \right)^2-4\La^6 =
\left(x^2 +\eta \La^2 \right)^2\left( x^2+4 \eta \La^2 \right)  \nonu
\end{eqnarray}
where $\eta$ satisfies $\eta^3=-1$ or equivalently $\eta^6=1$.
Recall that the matrix model curve for pure $SO(5)$ gauge theory
\cite{ao2} is identical to this curve. 
That is, $N-N_f=5$.
The factorization problem \cite{afo1}
resulted in the matrix model curve 
$
\widehat{y_m}^2 = x^2  \left( x^2- \al^2 \right)^2 -
4  \La^6 
$. 
There exists a symmetry breaking $SO(6) \to \widehat{SO(4)} 
\times U(1) $.
For $\al^2=-3\eta \La^2$, this can be factorized as $
\left(x^2+4\eta \La^2 \right) \left(x^2+\eta \La^2 \right)^2
$. 
For $\al^2=2\La^2$, the curve $x^2 
\left( x^2- \al^2 \right)^2 -\al^4x^2 $ 
will be
$
x^4 \left( x^2 - 4 \La^2  \right)
$. The latter 
implies the degenerated case. 
The characteristic function $P_6(x)= x^4 \left( x^2+3\eta \La^2 \right)$
becomes $2 x^3\rho^3 \La^3 {\cal T}_3\left( \frac{x}{2\rho \La}\right)$.
These are the vacua surviving when the ${\cal N}=2$ theory
is perturbed by a quadratic ($n=0$) superpotential and the 
$SO(6)$ gauge theory becomes massive at low energies.

Let us demonstrate how one can obtain the precise values for
$(N_0,N_1)$. 
The results for a matrix model curve and a characteristic function 
can be rewritten as follows (we put $\La=1$ and $\eta=-1$): 
\begin{eqnarray}
\widehat{y_m}^2=
\left(x^2-4 \right) \left(x^2-1 \right)^2,
\qquad  P_6(x)=
x^4 \left(x^2 -3 \right). 
\nonu
\end{eqnarray}
By using these relations, the function $T(x)=\mbox{Tr} \frac{1}{x-\Phi}$ 
is given as \cite{afo1} 
\begin{eqnarray}
T(x)=\frac{P_6^{\prime}(x)-\frac{(N_f+2)}{x}P_6(x)}{\sqrt{P_6(x)^2-
4x^6\La^{6}}}+\frac{(N_f+2)}{x}=-\frac{3}{\sqrt{x^2-4}}+\frac{3}{x}.
\nonu
\end{eqnarray}
There 
exist three branch cuts on the $x$-plane $[-2, -1-\ep],
[-1, 1]$, and $[1+\ep, 2]$ before taking the limit $x_1 \to x_0 
(\ep \to 0)$. \footnote{Recall that a matrix model curve is given by 
$y_m^2= \prod_{i=0}^{2} \left(x^2+ x_i^2 \right)$.} Since we are assuming 
$n=0$ and $n=1$ singular case, these three-branch cuts 
are joined at the 
locations of $x= \pm 
1$ after taking the limit and they reduce to a single-branch cut $[-2,2]$. 
Therefore, we can explicitly calculate the values $(N_0,N_1)$ as 
follows:
\begin{eqnarray}
N_0&=& \frac{1}{2\pi i} \oint_{A_0} T(x) dx=
 \frac{2}{2\pi i}\int_{-1}^{1}\left
(-\frac{3}{\sqrt{x^2-4}}+\frac{3}{x} \right)dx=\left(\frac{6}
{\pi}\int_0^{1}\frac{dx}{\sqrt{4-x^2}}\right)+3 =4, \nonu 
\\
N_1&=&
 \frac{1}{2\pi i} \oint_{A_1} T(x) dx=
\frac{2}{2\pi i}\int_{1}^{2}\left(-\frac{3}
{\sqrt{x^2-4}}+\frac{3}{x} \right)dx=\frac{3}{\pi} 
\int_{1}^{2}\frac{dx}{\sqrt{4-x^2}}=1
\nonu
\end{eqnarray}
as we expected.
\footnote{
Recently \cite{jan2003}, 
the construction of $T(x)$ on an elliptic curve in terms of 
three discrete parameters and two continuous ones was studied in the 
breaking pattern $U(N) \to U(N_1) \times U(N_2)$ where $N=N_1+ N_2$.
This method will give some hints to obtain this 1-form 
explicitly without resort to a characteristic function we did here.}

$\bullet$ $SO(6)$ with $N_f=2$

Once again 
the factorization problem \cite{afo1}
resulted in the matrix model curve 
that contains an overall $x^2$ factor
implying a degenerated case.

$\bullet$ $SO(6)$ with $N_f=3$

The factorization problem \cite{afo1}
turned out 
the matrix model curve that has an overall $x^2$ factor 
and is classified as a degenerated case
which will be studied later.

$\bullet$ $SO(7)$ with $N_f \geq 1$

The curve for $SO(7)$ with $N_f$ flavors is 
the same as $SO(6)$ curve with $(N_f-1)$ flavors.
In particular, since the SW curve of $SO(7)$ theory  with $N_f=1$ 
is identical to the $SO(6)$ gauge theory with $N_f=0$, 
the factorization problem in \cite{ao} implies 
that the matrix model curve is
\bea
y_m^2=x^2\left(x^2+ 4 \eta \La^2  \right)^2 + 4 \eta^2 \La^4 x^2 +
16 \eta^3 \La^6 =
\left(x^2 + 2\eta \La^2 \right)^2 
\left( x^2+4 \eta \La^2 \right), \qquad \eta^4=1  
\nonu
\eea
where the breaking pattern $SO(7) \to \widehat{SO(5)} \times U(1)$ 
exists.
\footnote{
Let us check how one can determine the precise values for
$(N_0,N_1)$. 
The results for a matrix model curve and a characteristic function are
given by (we put $\La=1$ and $\eta=-1$): 
$
\widehat{y_m}^2=
\left(x^2-4 \right) \left(x^2-2 \right)^2,
  P_6(x)=
x^2 \left(x^4-4x^2 +2 \right)$. 
By using these relations, the function $T(x)$ 
is given by
$
T(x)=\frac{P_6^{\prime}(x)-\frac{(N_f+1)}{x}P_6(x)}{\sqrt{P_6(x)^2-
4x^4\La^{8}}}+\frac{(N_f+2)}{x}=-\frac{4}{\sqrt{x^2-4}}+\frac{3}{x}$.
Note that there was a typo in $T(x)$ \cite{ao} and there should be
present an extra $1/x$ term in $T(x)$. 
There 
exist three-branch cuts on the $x$-plane $[-2, -\sqrt{2}-\ep],
[-\sqrt{2}, \sqrt{2}]$, and $[\sqrt{2}+\ep, 2]$ 
before taking the limit $x_1 \to x_0 
(\ep \to 0)$. These three-branch cuts 
are joined at the 
locations of $x= \pm 
\sqrt{2}$ 
after taking the limit and they reduce to a single-branch cut $[-2,2]$. 
Therefore, we can explicitly compute the values $(N_0,N_1)$ as 
follows:
$
N_0= 
 \frac{2}{2\pi i}\int_{-\sqrt{2}}^{\sqrt{2}}\left
(-\frac{4}{\sqrt{x^2-4}}+\frac{3}{x} \right)dx=5,
N_1=
\frac{2}{2\pi i}\int_{\sqrt{2}}^{2}\left(-\frac{4}
{\sqrt{x^2-4}}+\frac{3}{x} \right)dx=1$.  }
At the intersections with $n=0$ branch, the characteristic function 
$P_6(x)$ can be written in terms of the Chebyshev polynomial of degree 4
with an appropriate identification between these two functions. 

$\bullet$ $SO(8)$ with $N_f \geq 1$

The curve for $SO(8)$ with $N_f=1$ can be obtained from the factorization 
problem. The result is, by applying the method of \cite{afo1} to the 
present case here,
$
\widehat{y_m}^2 = \left( x^2 + \frac{3 \pm \sqrt{5}}{2} \eta 
\La^2 \right)^2 \left( x^2 + 4 \eta \La^2 \right)
$ with $\eta^5=-1$.
This curve is exactly the same as the expression from the glueball approach 
by inserting the values $(N_0,N_1,N_f)=(6,1,1)$ into (\ref{curve}) where
$SO(8) \to \widehat{SO(6)} \times U(1)$.
Note that the above matrix model curve is the same as the curve for 
pure $SO(7)$ gauge theory \cite{ao2}:
$N-N_f=7$. 
Although there is no information about the curve with larger flavors from 
factorization problem, one 
predicts a matrix model curve through (\ref{curve}). 

$\bullet$ $SO(9)$ with $N_f \geq 1$

The curve for $SO(9)$ with $N_f$ flavors is 
the same as $SO(8)$ curve with $(N_f-1)$ flavors.
In particular, since the SW curve of $SO(9)$ theory  with $N_f=1$ 
is identical to the $SO(8)$ gauge theory with $N_f=0$, 
the factorization problem in \cite{ao} implies 
that the matrix model curve becomes
\bea
y_m^2=x^2\left(x^2+ 5 \eta \La^2  \right)^2 + 8 \eta^2 \La^4 x^2 +
36 \eta^3 \La^6 =
\left(x^2 + 3\eta \La^2 \right)^2 
\left( x^2+4 \eta \La^2 \right), \qquad
\eta^6=1
\nonu
\eea
where the breaking pattern $SO(9) \to \widehat{SO(7)} \times U(1)$ 
exists.
\footnote{
By using
$
\widehat{y_m}^2=
\left(x^2-4 \right) \left(x^2-3 \right)^2,
  P_8(x)=
x^4 \left(x^2-3 \right)^2 -2x^2 $,
the function $T(x)$ 
is given by
$
T(x)=\frac{P_8^{\prime}(x)-\frac{(N_f+1)}{x}P_8(x)}{\sqrt{P_8(x)^2-
4x^4\La^{8}}}+\frac{(N_f+2)}{x}=-\frac{6}{\sqrt{x^2-4}}+\frac{3}{x}$.
There 
exist three-branch cuts on the $x$-plane $[-2, -\sqrt{3}-\ep],
[-\sqrt{3}, \sqrt{3}]$, and $[\sqrt{3}+\ep, 2]$ 
before taking the limit $x_1 \to x_0 
(\ep \to 0)$. These three-branch cuts 
are joined at the 
locations of $x= \pm 
\sqrt{3}$ 
after taking the limit and they reduce to a single-branch cut $[-2,2]$. 
Therefore, we can explicitly compute the values $(N_0,N_1)$ as 
follows:
$
N_0= 
 \frac{2}{2\pi i}\int_{-\sqrt{3}}^{\sqrt{3}}\left
(-\frac{6}{\sqrt{x^2-4}}+\frac{3}{x} \right)dx=7,
N_1=
\frac{2}{2\pi i}\int_{\sqrt{3}}^{2}\left(-\frac{6}
{\sqrt{x^2-4}}+\frac{3}{x} \right)dx=1$.}
At the intersections with $n=0$ branch, the characteristic function 
$P_8(x)$ can be written in terms of the Chebyshev polynomial of degree 6
with an appropriate identification between these two functions.
These are the vacua that survive when the ${\cal N}=2$ theory is perturbed
by a quadratic superpotential ($n=0$).



Recalling the replacement of $(N_0-2)$ in pure gauge theory 
result with $(N_0-N_f-2)$ for massless flavors 
we can get the coupling constant near the singularity without explicit 
calculation,
\begin{eqnarray}
\frac{1}{2\pi i}\tau_{ij}=\frac{\partial^2 
{\cal F}_p(S_k)}{\partial S_i \partial S_j}-\delta_{ij}
\frac{1}{N_i}\sum_{l=1}^n N_l \frac{\partial^2 {\cal F}_p(S_k)}
{\partial S_i \partial S_l}-\delta_{ij} \left(\frac{N_0 -N_f- 2}{N_i}
\right)
\frac{\partial^2 {\cal F}_p(S_k)}{\partial S_0\partial S_i},
\nonu
\end{eqnarray}
where  $i, j=1,2, \cdots, n$.
In the quartic tree-level superpotential case ($n=1$) we are considering, 
there is only one coupling constant and it is
given by, due to the cancellation of first two
terms above, 
\begin{eqnarray}
\frac{1}{2\pi i}\tau=
- \frac{i \pi}{16}  \frac{\left(N_0-N_f-2 \right)^2}{ 
\left(N-N_f-2 \right)^2} 
\frac{1}{ \log \left(\frac{16}{1-k^{\prime 2}}\right)},
\nonu
\eea
where $ 
k^{\prime ^2}=
\frac{x_0^2\left(x_2^2-x_1^2\right)}{x_1^2\left(x_2^2-x_0^2\right)}
$.
Since we are assuming asymptotically free gauge theory, the 
quantity $(N-N_f-2)$ is greater
 than zero. When the condition $(N_0-N_f-2)=0$ is satisfied, $\tau$ 
becomes zero. In this case, the asymptotic 
freedom for $SO(N_0)$ gauge theory breaks down. 
Therefore 
the situation is the same as the one-cut case (equivalently $n=0$ case) 
in which the coupling constant is trivially zero. Our result gives a 
consistency.

\section{The curve for degenerated $SO(N)$ gauge theory
with massless
flavors}
\setcounter{equation}{0}

\indent

In the non-degenerated case, every root of $W_3^{\prime}(x)$
possess at least one D5 brane wrapping around it. 
For the degenerated case,
some roots do not contain wrapping D5 branes around them. 
At the classical limits, there are symmetry breakings characterized by
\bea
SO(N) \to \widehat{SO(N_0)} 
\times U(N_1), \qquad SO(N) \to U([N/2]), \qquad SO(N) 
\to \widehat{SO(N)}.
\nonu
\eea
The last pattern occurs only if further constraints    
are imposed on the first 
two breaking patterns. 
For degenerated  case we can write the matrix model curve which is 
valid on the whole degenerated branch. 
As we did for pure gauge theory, the dual periods are defined by
the integrals over $x$ and the equation of motion of a single field
$F$ provides
one relation.
Recalling the simple 
replacements $(N-2) \to (N-N_f-2)$ and $(N_0-2) \to (N_0-N_f-2)$ 
we made before, 
we can obtain the matrix model curve 
\footnote{For the degenerated case also, the relation between 
$W_3^{\prime}(x)$ and $F_4(x)$ can be written as
\bea
y_{m,d}^2 &= & F_4(x) = \left(\frac{W_3^{\prime}(x)}{x} \right)^2+
4F, \ \qquad \qquad \qquad
N_f< N-3, \nonu \\
y_{m,d}^2 & = & F_4(x) = \left(\frac{W_3^{\prime}(x)}{x} \right)^2+
4F - 4x^2\Lambda^{2}, \ \ \qquad
N_f= N-3. 
\nonu
\eea
In particular, when $N_f=N-3$, the matrix model curve 
$y_{m,d}^2$ has an extra contribution  
from $4x^2 \La^2$. Therefore in this particular case, the curve should be
$y_{m,d}^2= \left(x^2+m- 2\La^2 \right)^2+4F$ and $m-2\La^2=
\frac{K^2-F}{K}$.
\label{parti}} 
as 
\begin{eqnarray}
y_{m,d}^2=\left(x^2+m \right)^2+4F,\qquad \ m=\frac{K^2-F}{K},
\qquad K\equiv \left[\frac{(-\La^2)^{N-N_f-2}}{(-F)^{N_1}} 
\right]^{\frac{1}{N_0-N_f-2}}. 
\label{formula}
\end{eqnarray}
Here we rescaled the function $K$ and corrected a typo appeared 
in \cite{ao2}.
This formula depends on the parameter $F, N_0, N_1$, and $N_f$.
Turning on a field $F$ to the particular values will 
reduce to the last breaking pattern $SO(N) 
\to \widehat{SO(N)}$ above.
When $(N_0-N_f-2)=0$ (we will meet this situation in the examples 
below), the above formula becomes ill-defined. We  go back 
to the derivation of the formula 
\footnote{In this case, there is a relation 
$2N_1 \log \big| \frac{4\La^2}{b^2-a^2} \big|$=0. By putting $\eta=
\pm \frac{4\La^2}{b^2-a^2}$ and realizing $D=\frac{1}{2} 
\left( a^2 +b^2 \right)$, one gets the above formula.}
and can find another formula 
for this special case,
\begin{eqnarray}
y_{m,d}^2=\left(x^2+D \right)^2-\frac{4\La^4}{\eta^2},\qquad
 \eta^{2N_1}=1 ,\qquad \mbox{for} \qquad (N_0-N_f-2)=0 
\label{formu}
\end{eqnarray}
where $D$ is arbitrary parameter.

To demonstrate these general solutions 
(\ref{formula}) and (\ref{formu}), 
let us consider $SO(N)$ 
gauge theories with $N_f (\le N-3)$ flavors where $N=4,5$ or $6$.

$\bullet$ $SO(4)$ with  $N_f=1$

There are two breaking patterns on the degenerated case: 
$\widehat{SO(2)}\times U(1)$ and $U(2)$. 

1) For the breaking pattern $SO(4) \to 
\widehat{SO(2)}\times U(1)$, by substituting the values
$(N,N_1,N_f)=(4,1,1)$ into (\ref{formula}) with the footnote
\ref{parti} 
(recall that 
$N=N_0+2N_1$),  
the matrix model curve from glueball approach 
can be represented by
\begin{eqnarray}
y_{m,d}^2=\left(x^2+m - 2\La^2 \right)^2+4F,\qquad  m=
\frac{F}{\La^2}+\La^2. \nonu 
\end{eqnarray}
We find that the solutions given in \cite{afo1} perfectly 
agree with the above result for $y_{m,d}^2$ from glueball approach. 
In other words, the factorization 
problem characterized by 
$P^2_4(x)-4x^6 \La^2=H^2_2(x) \left[ \left(x^2-D 
\right)^2+4F \right]$ 
reproduces 
the above solution where $D=\La^2-\frac{F}{\La^2}$ (For the notation of
\cite{afo1}, $a$ is equal to $-2D$ and $\left(\al^2+  2\La^2 \right)$ 
corresponds to $\La^2 -
\frac{F}{\La^2}=D$). 
Then $\al^2=-\La^2 -\frac{F}{\La^2}$ which is equal to $-m$.
For $F=0$, there exists a Special branch with 
$SO(4) \to \widehat{SO(4)}$ and for $ \left(m-2\La^2 \right)^2 +4F=0$, 
there exists a Chebyshev branch 
with $SO(4) \to \widehat{SO(4)}$.

2) For the breaking pattern $SO(4) \to U(2)$,
we obtain the following matrix model curve from the glueball approach,
by plugging the values $(N,N_1,N_f)=(4,2,1)$ into (\ref{formula}) with
the footnote \ref{parti},
\begin{eqnarray}
y_{m,d}^2=\left(x^2+m -2\La^2 \right)^2+4F,\qquad  m=
\left(-\frac{F^2}{\La^2} \right)^{\frac{1}{3}}-F\left(-
\frac{\La^2}{F^2} \right)^{\frac{1}{3}} + 2\La^2. \nonu 
\end{eqnarray}
Again by using the different parametrization (as in previous 
case) from the one in \cite{afo1}, 
we can obtain the same matrix model curve with $D$ which is
equal to $-m+2\La^2$ above 
where the relation 
$6 \al^2 \La^2 + 14 \La^4 - 2 \left( \al^2 
+ 3\La^2 \right) \sqrt{4 \al^2 \La^2 + 9\La^4}=4F$ (corresponding to 
$b-a^2/4$ in the notation of \cite{afo1}) will provide 
one solution for $\al^2=-m$ in the above. 
Therefore, the results obtained from two approaches 
are equivalent to 
each other.

$\bullet$ $SO(5)$ with  $N_f=1$

There are two breaking patterns on the degenerated case: 
$\widehat{SO(3)}\times U(1)$ and $\widehat{SO(1)} \times U(2)$. 

1) For the former 
case $SO(5) \to \widehat{SO(3)}\times U(1)
$, since the condition $(N_0-N_f-2)=0$ is satisfied, 
the matrix model curve can be read off,
from (\ref{formu}),
\begin{eqnarray}
y_{m,d}^2=\left(x^2+D \right)^2-4\La^4. \nonu
\end{eqnarray}
This is exactly the first kind of solution given in \cite{afo1} and
the $D$ here corresponds to $a/2=-\al^2$ there and 
$-4\La^4$ corresponds to 
$(b-a^2/4)$. For $D^2=4\La^4$, there is a Chebyshev vacuum where
$SO(5) \to \widehat{SO(5)}$.

2) For the second breaking pattern $SO(5) \to 
\widehat{SO(1)} \times U(2)$,
by paying attention 
to the value of $N_0$, 
we simply put $N_0=1$ and find the following 
matrix model curve,
\begin{eqnarray}
y_{m,d}^2=\left(x^2+m \right)^2+4F,\qquad  
m=\frac{\eta F}{\La^2}-\eta \La^2 \nonu 
\end{eqnarray}
where $\eta$ is $2$-nd root of unity. These solutions are 
exactly the same as the ones  from factorization 
problem in \cite{afo1} ($m=a/2$ and $b-a^2/4=4F$).
For $a=\pm 2 \La^2$ ($+$ sign corresponds to 
$\eta =-1$ above and $-$ sign corresponds to $\eta=1$), 
there is a Special vacuum where $
SO(5) \to \widehat{SO(5)}$.
This curve is the same as the matrix model curve for pure
degenerated $SO(4)$ gauge theory \cite{ao2} ($N-N_f=4$) 
where $SO(4) \to U(2)$.

$\bullet$ $SO(6)$ with  $N_f=1$

In this case, there are three kinds of breaking patterns on the 
degenerated case: $\widehat{SO(4)}\times U(1)$, 
 $\widehat{SO(2)}\times U(2)$,
and $U(3)$.

1) For the breaking pattern $SO(6) \to \widehat{SO(4)}\times U(1)$, 
the matrix model curve is represented 
as from (\ref{formula}) with $(N,N_1,N_f)=(6,1,1)$ 
\begin{eqnarray}
y_{m,d}^2=\left(x^2+m \right)^2+4F,\qquad  m=\frac{\La^6}
{F}-\frac{F^2}{\La^6}. \nonu 
\end{eqnarray}
Identifying $4F$ here with $b$ in \cite{afo1} (and $m=-a$),
we can see the exact 
agreement.
Recall that this curve is identical to the matrix model curve 
for pure degenerated $SO(5)$ gauge theory \cite{ao2} ($N-N_f=5$) where 
$SO(5) \to SO(3) \times U(1)$.

2) On the other hand, for the breaking pattern $SO(6)\to U(3)$,
 the 
matrix model curve can be written as from (\ref{formula}) 
together with $(N,N_1,N_f)=(6,3,1)$
\begin{eqnarray}
y_{m,d}^2=\left(x^2+m \right)^2+4F,\qquad  m=\frac{\eta F}
{\La^2}-\eta^2\La^2 \nonu 
\end{eqnarray}
where $\eta^3=1$. By identifying $(a,b)$ with $(-m,4F)$,
we can see the complete agreement with the first, second and 
third solutions of the equation $(6.38)$ in 
\cite{afo1} although we have not written them explicitly.
For $b=4F=0$, this  leads to  a Special vacuum of 
$SO(6) \to \widehat{SO(6)}$.

3) For the breaking pattern $SO(6) \to \widehat{SO(2)} 
\times U(2)$, the solution from factorization was not new one but
can be written as the previous solution we have discussed already.  

$\bullet$ $SO(6)$ with  $N_f=2$

1) For the breaking pattern $SO(6) \to
\widehat{SO(4)}\times U(1)$, the 
matrix model curve can be written as from (\ref{formu}) 
\begin{eqnarray}
y_{m,d}^2=\left(x^2+D \right)^2-4\La^4. \nonu
\end{eqnarray}
This corresponds to the first kind solution in \cite{afo1}. 
When $D^2=4\La^4$, this will lead to a Chebyshev vacuum with
$SO(6) \to \widehat{SO(6)}$.

2) For the breaking pattern $SO(6) \to
\widehat{SO(2)}\times U(2)$,
 the matrix 
model curve is given by from (\ref{formula}) with $(N,N_1,N_f)=(6,2,2)$
\begin{eqnarray}
y_{m,d}^2=\left(x^2+m \right)^2+4F,\qquad  
m=\frac{\eta F}{\La^2}-\eta \La^2 \nonu 
\end{eqnarray}
where $\eta$ is $2$-nd root of unity. 
This can be seen from the degenerated pure $SO(4)$ gauge theory with 
the breaking pattern $SO(4) \to U(2)$ \cite{ao2} 
through the addition map.
If we change the 
parametrization by $4F=4\eta a \La^2+4\La^4$, the above matrix 
model curve can be rewritten as
\begin{eqnarray}
y_{m,d}^2=\left(x^2+a \right)^2+4\eta\La^2 x^2, 
\nonu
\end{eqnarray}
which is exactly the same as the second kind of solution 
in \cite{afo1}. For $a=0$, it gives a Chebyshev vacuum of
$SO(6) \to \widehat{SO(6)}$. For $a=-\eta \La^2$, a Special 
branch appears. 
\footnote{
For the breaking pattern $SO(6) \to U(3)$,
we can predict the following matrix 
model curve, from (\ref{formula}) with $(N,N_1,N_f)=(6,3,2)$,
$
y_{m,d}^2=\left(x^2+m \right)^2+4F$, and $  m=\frac{\eta 
F^{\frac{3}{4}}}{\La}-\frac{1}{\eta}\La F^{\frac{1}{4}} 
$
where $\eta^8=1$. 
In this case, the factorization 
problem becomes very complicated. 
Since we cannot obtain the explicit results 
for the matrix model curve from the factorization, 
we have not checked them in this example.}

$\bullet$ $SO(6)$ with  $N_f=3$

In this case, there are three breaking patterns: 
$\widehat{SO(4)}\times U(1)$, $\widehat{SO(2)}\times U(2)$ 
and $U(3)$. 

1) For the first two cases, we expect that the matrix 
model curves are the same as the ones in $SO(4)$ theory with $N_f=1$ 
as discussed above, which comes from the addition map for massless case.
For the breaking pattern $SO(6) \to 
\widehat{SO(4)}\times U(1)$, the 
matrix model curve can be represented as from (\ref{formula})
with the footnote \ref{parti}
\begin{eqnarray}
y_{m,d}^2=\left(x^2+m - 2\La^2 \right)^2+4F,\qquad  m=
\frac{F}{\La^2}+\La^2.
 \nonu
\end{eqnarray}
From the result of 
\cite{afo1}, the matrix model curve was $\widehat{y_m}^2= \left(x^2 -
 \al^2 -2\La^2\right)^2-4\al^2 -4\La^4$. By introducing 
a new field $4F= -4\al^2 \La^2 -4\La^4$,  one can identify
$\al^2=-\La^2 -\frac{F}{\La^2}$ which is equal to $-m$ in the above.

2) For the breaking pattern, $SO(6) \to 
\widehat{SO(2)}\times U(2)$, from (\ref{formula}) with the footnote 
\ref{parti} the matrix model curve 
can be obtained, with
$(N,N_1,N_f)=(6,2,3)$,
\begin{eqnarray}
y_{m,d}^2=\left(x^2+m -2\La^2 \right)^2+4F,\qquad  
m=\left(- \frac{F^2}{\La^2}\right)^{\frac{1}{3}}-
F\left(- \frac{\La^2}{F^2} \right)^{\frac{1}{3}} + 2\La^2. \nonu 
\end{eqnarray}
As expected, this matrix model curve is the same as 
the one in $SO(4)$ theory with $N_f=1$ case
and agrees with the one from 
the factorization problem. 
In other words, the factorization problem $P_6^2(x)-4x^4\La^4=
x^4 H_2^2(x)\left[\left(x^2-D \right)^2+4F \right]$ 
reproduces the above 
solution where $D=-\left(- \frac{F^2}{\La^2}\right)^{\frac{1}{3}}+
F\left(- \frac{\La^2}{F^2} \right)^{\frac{1}{3}}$.
When $F=0$, it gives a Special vacuum of $SO(6) \to 
\widehat{SO(6)}$. For $\left(m-2 \La^2\right)^2 +4F=0$, 
it leads to a Chebyshev vacuum
of $SO(6) \to \widehat{SO(6)}$.
\footnote{
The last one is the $SO(6) \to U(3)$ breaking pattern. 
In this case the matrix 
model curve is given by, from 
(\ref{formula}) and the footnote \ref{parti} 
with $(N,N_1,N_f)=(6,3,3)$, 
$
y_{m,d}^2=\left(x^2+m -2\La^2 \right)^2+4F$, and $  
m=\left(\frac{F^3}{\La^2} \right)^{\frac{1}{5}}-
F\left(\frac{\La^2}{F^3} \right)^{\frac{1}{5}} + 
2\La^2$.
Since the
factorization problem becomes very complicated,
we have not checked them in this example.}

In summary, for the  $SO(N)$ 
gauge theories with $N_f (\le N-3)$ flavors where $N=4,5$ or $6$,
we have convinced ourselves that the solutions 
(\ref{formula}) and (\ref{formu}) coincide with the matrix model curve
from the strong-coupling approach except that we could not check the 
$SO(6)$ gauge theory with $N_f=2,3$ where there exists a breaking pattern 
$SO(6) \to U(3)$ (in this case, $N_0=0$). 
The main obstacle why 
the factorization problem for $SO(N) \to U([N/2])$ 
becomes so complicated was the SW curve
does not possess any power of $x^2$, contrary to the factorization problem 
for $SO(N) \to \widehat{SO(N_0)} \times U(N_1), N_0 \neq 0$
in which the SW curve has a power of $x^2$ (either Chebyshev or Special
branch) making the factorization easier.  
Therefore, at least one expects that our matrix model curve 
(\ref{formula}) and (\ref{formu}) will predict 
the solutions for the  degenerated $SO(N)$ 
gauge theories with $N_f (\le N-3)$ massless flavors where $N \ge 7$
and $SO(N) \to \widehat{SO(N_0)} \times U(N_1), N_0 \neq 0$.

\section{The $n=0$ and $n=1$ singularity: $USp(2N)$ gauge theory
with massless flavors
 \label{USpmassless} }
\setcounter{equation}{0}


\indent

As in the $SO(N)$ case discussed previous section, 
we can obtain the matrix model curve 
without explicit calculation by looking at the pure $USp(2N)$ gauge theory. 
We  only have to replace the expression $(2N+2)$ in 
the result of \cite{ao2} for the matrix model curve with 
$(2N-N_f+2)$. With this replacement 
in mind, we immediately obtain the 
matrix model curve near the singularity on the non-degenerated case
from the glueball approach 
\footnote{The effective glueball superpotential for massless flavors 
is
$
W_{\rm eff}=2\pi i \left[ (2N_0+2-N_f)\Pi_0+
2N_1\Pi_1 \right]-2(2N+2-N_f) S \log \left( \frac{\La}
{\La_0} \right).$}: 
\begin{eqnarray}
y_m^2&=&x^2 \left(x^2+m \right)^2+
\langle f_2 \rangle x^2 +\langle f_0 \rangle =
\left[ x^2 + 2 \eta \La^2 \left(1+c \right) \right]^2 \left( x^2 +
4 \eta \La^2 \right), \nonu
\end{eqnarray}
where the fields, parameter of a superpotential, 
and glueball field are given by 
\begin{eqnarray}
\langle f_2 \rangle &=& 4\eta^2 \La^4 \left(1+2 c 
\right),\qquad 
\langle f_0 \rangle=16\eta^3 \La^6 \left(1+c 
\right)^2, \nonu \\
m &=& 2\eta \La^2 \left(2+ c \right), \qquad  
\langle S \rangle=-\eta^2\La^4 \left(1+2 c 
\right), \qquad c \equiv \cos \left( \frac{2\pi N_1}{2N-N_f+2} \right).   
\label{curve1}
\end{eqnarray}
Here $\eta$ is the $(2N-N_f+2)$-th root of unity, that is 
$\eta^{2N-N_f+2}=1$. 
\footnote{
By recognizing the matrix model curve for $USp(2N)$ gauge theory
characterized by $y_m^2 = \prod_{i=0}^{2} \left(x^2-x_i^2\right)$,
one can read off the 
matrix model curve for $USp(2N)$ with flavors from the 
curve for $SO(N)$ gauge theory with flavors by replacing $(N-2) \to 
(2N+2), x_i^2 \to -x_i^2$ where $i=0,1,2$
as follows:
$
x_2^2=-4\eta \La^2$, and $x_0^2=-2\eta \La^2
\left(1+c \right)
$
where $\eta$ satisfies the condition
$
\eta^{2N-N_f+2}=1$ for $(2N-N_f+2)$ 
even and $ \eta^{2N-N_f+2}=-1$  
for $(2N-N_f+2)$ odd.
When we define $\epsilon \equiv -\eta$ 
it satisfies $\epsilon^{2N-N_f+2}=1$ for both cases. 
The solutions can be rewritten as
$
x_2^2=4\epsilon \La^2$, $ x_0^2=2\epsilon 
\La^2\left(1+c 
\right)$, and $\epsilon^{2N-N_f+2}=1
$ where $c$ is given in (\ref{curve1}). This is exactly the same as 
(\ref{curve1}) by using the different parametrization.} 
This matrix model curve with massless flavors is exactly the same as pure case 
with the replacement $(2N+2) \to (2N-N_f+2)$. 
Note that the above solutions have the special combination $(2N-N_f+2)$ 
which reflects the addition map discussed in \cite{afo2}, relating the 
Chebyshev branches (or the Special branches) of two different gauge groups. 
Namely, the curve of $USp(2N)$ with $N_f$ massless flavors can be 
reduced to the curve of $USp(2N-2r)$ with 
$(N_f-2r)$ massless flavors.  

Let us demonstrate these general features (\ref{curve1}) 
explicitly by comparing 
them with the results from strong-coupling approach 
obtained in \cite{afo2} already. Let us consider $USp(2N)$ 
gauge theories with $N_f (\le 2N+1)$ flavors where $N=2$ or 3.

$\bullet$ $USp(4)$ with $N_f= 1$

From the results in \cite{afo2},
the matrix model curve has an extra double root at 
$d=\frac{3\pm \sqrt{5}}{2} \La^2 \ep$ on 
the non-degenerated case. It was 
factorized as $\widehat{y_m}^2= \left(t+\frac{3\pm \sqrt{5}}{2}\epsilon
 \La^2 \right)^2 \left(t+4\ep \La^2 \right)$ 
where $\epsilon$ is $5$-th root of unity.
\footnote{We use a relation $t=x^2$ all the time in this paper. }
On the other hand, the matrix model curve from the glueball approach
becomes, from (\ref{curve1}) with $(N,N_1,N_f)=(2,1,1)$,
\bea
y_m^2 =\left(x^2 + \frac{3\pm \sqrt{5} }{2} \eta \La^2  \right)^2 
\left( x^2 + 4 \eta \La^2\right)
\nonu
\eea
where the other values in different parametrization are given by 
\begin{eqnarray}
m=\frac{\left(7\pm \sqrt{5} \right)}{2} \eta \La^2,\qquad \langle f_2 
\rangle=2 \left(1\pm \sqrt{5} \right) \eta^2 \La^4,\qquad 
\langle f_0 \rangle=
2 \left(7\pm 3\sqrt{5} \right) \eta^3\La^6.
 \nonu
\end{eqnarray}
Here $\eta$ is $5$-th root of unity. Now it is easy to see 
the exact agreement between two results.

$\bullet$ $USp(4)$ with $N_f= 2$

The matrix model curve obtained from 
the factorization problem can be represented, at singular point, as 
$\widehat{y_m}^2=\left(t+2\eta \La^2 \right)^2 \left(t+4\eta 
\La^2 \right)$ 
where 
$\eta$ is $4$-th root of unity (Note that there is a typo in 
\cite{afo2}). 
The characteristic function
$B_6(x)=x^2 \left( x^4 + 4 \eta \La^2 x^2  + 2  \eta^2 \La^4 \right)$ 
will become 
$2 \rho^2 x^2 \La^4 {\cal T}_2 
\left( \frac{x^2}{2 \rho \La^2}+ 1 \right)$ 
with $\rho^4=1$ by identifying $\rho=\eta$. 
These are the vacua surviving when the ${\cal N}=2$ theory
is perturbed by a quadratic ($n=0$) superpotential and the 
$USp(4)$ gauge theory becomes massive at low energies.
To see the correctness of our formula (\ref{curve1}) 
we put simply 
the values $(2N_0,N_1,N_f)=(2,1,2)$ (Note that $2N=2N_0+2N_1$) 
where $USp(4) \to \widehat{USp(2)}
\times U(1)$ 
into the formula (\ref{curve1}) and get the 
following matrix model curve,
\begin{eqnarray}
y^2_m=x^2\left(x^2+4\eta \La^2 \right)^2+4 \eta^2 \La^4 x^2 +16\eta^3 
\La^6 = \left(x^2 + 2\eta \La^2 \right)^2 \left(x^2 +4 \eta \La^2 \right). 
\nonu
\end{eqnarray}
Now there exists a perfect agreement. 

Let us demonstrate how one can determine the precise values for
$(2N_0,N_1)$. 
The results can be rewritten as follows (we put $\La=1$ and $\eta=1$): 
\begin{eqnarray}
\widehat{y_m}^2=
\left(x^2+4 \right) \left(x^2+2 \right)^2,
\qquad  B_6(x)=
x^2 \left(x^4 +4x^2+2 \right). 
\nonu
\end{eqnarray}
By using these relations, the function $T(x)=\mbox{Tr} \frac{1}{x-\Phi}$ 
is given as \cite{afo2} 
\begin{eqnarray}
T(x)=\frac{B_6^{\prime}(x)-\frac{N_f}{x}B_6(x)}{\sqrt{B_6(x)^2-
4x^4 \La^{8}}}+\frac{(N_f-2)}{x}=\frac{4}{
\sqrt{x^2+4}}.
\nonu
\end{eqnarray}
There 
exist three branch cuts on the $x$-plane $[-2i, -(\sqrt{2}+\ep)i],
[-\sqrt{2}i, \sqrt{2}i]$, and $[(\sqrt{2}+\ep)i , 2i]$ 
before taking the limit $x_1 \to x_0$. 
Since we are assuming 
$n=0$ and $n=1$ singular case, these three-branch cuts 
are joined at the 
locations of $x= \pm 
\sqrt{2}i$ 
after taking the limit and they reduce to a single-branch cut $[-2i,2i]$. 
Therefore, we can explicitly calculate the values $(2N_0,N_1)$ as 
follows:
\begin{eqnarray}
2N_0&=& \frac{1}{2\pi i} \oint_{A_0} T(x) dx=
 \frac{2}{2\pi i}\int_{-\sqrt{2}i}^{\sqrt{2}i}\frac{4}{
\sqrt{x^2+4}} dx =2, \nonu 
\\
N_1&=&
 \frac{1}{2\pi i} \oint_{A_1} T(x) dx=
\frac{2}{2\pi i}\int_{\sqrt{2}i}^{2i} \frac{4}{
\sqrt{x^2+4}} dx=1
\nonu
\end{eqnarray}
as we expected.

$\bullet$ $USp(4)$ with $N_f= 3$

The matrix model curve at the singularity was given in \cite{afo2} 
as $\widehat{y_m}^2= \left(t+\eta \La^2 \right)^2
\left(t+4\eta \La^2 \right)$ where $\eta^3=1$. 
Note that $\al^2$ corresponds to $-3 \eta \La^2$.
The $B_6(x)$ is equal to
$2 \rho^3 x^3 \La^3 {\cal T}_3 \left(\frac{x}{2\rho \La} \right)$
by putting $\rho=\eta$.
At these intersection points, the vacua survive 
when the ${\cal N}=2$ theory
is perturbed by a quadratic ($n=0$) superpotential.
If we put the values $(2N_0,N_1,N_f)=(2,1,3)$
where
 $USp(4) \to \widehat{USp(2)}
\times U(1)$  
into our formula (\ref{curve1}) 
we obtain the matrix 
model curve at the singularity,
\begin{eqnarray}
y^2_m=x^2\left(x^2+3\eta \La^2 \right)^2+4\eta^3 \La^6 =
\left( x^2 +\eta \La^2 \right)^2 \left( x^2 +4 \eta \La^2 \right). 
\nonu
\end{eqnarray}
We see the agreement between two approaches.

$\bullet$ $USp(4)$ with $N_f= 4$

There is no non-degenerated case and we will discuss it
in the degenerated case later.

$\bullet$ $USp(4)$ with $N_f= 5$

In \cite{afo2}, the factorization problem 
resulted in the matrix model curve $\widehat{y_m}^2$
which contains an overall factor $x^2$.
Therefore, there exists no solution for non-degenerated case.
We will come to this case for degenerated case later.

$\bullet$ $USp(6)$ with $N_f= 2$

1) For the first solution $USp(6) \to \widehat{USp(4)} \times
U(1)$ in \cite{afo2},
 the matrix model curve has an 
extra double root at $d=-\epsilon \La^2$, where $\epsilon$ is $6$-th 
root of unity. At the singular points, the matrix model can be factorized 
as $\widehat{y_{m}}^2=\left(t-3\epsilon 
\La^2 \right)^2 \left(t-4\epsilon \La^2 \right)$.
If we simply put the values $(2N_0,N_1,N_f)=(4,1,2)$ 
into our general formula 
(\ref{curve1}) for the
 matrix model curve at the singularities, 
we obtain the following results,
\begin{eqnarray}
y_{m}^2=x^2\left(x^2+5\eta \La^2 \right)^2+8\eta^2 
\La^4 x^2+36\eta^3\La^6=
\left(x^2 +3\eta \La^2 \right)^2 \left( x^2 + 4\eta \La^2 \right) 
\nonu
\end{eqnarray}
where $\eta$ is $6$-th root of unity. If we identify this $\eta$ with 
$-\epsilon$ in the solution for factorization problem, we see the 
agreement of two approaches. \footnote{
By using
$
\widehat{y_m}^2=
\left(x^2+4 \right) \left(x^2+3 \right)^2$, and $
B_8(x)=
x^2 \left(x^6 +6x^4+9x^2+2 \right)$, 
the function $T(x)$ 
is given by
$
T(x)=\frac{B_8^{\prime}(x)-\frac{N_f}{x}B_8(x)}{\sqrt{B_8(x)^2-
4x^4 \La^{8}}}+\frac{(N_f-2)}{x}=\frac{6}{
\sqrt{x^2+4}}$.
There 
exist three branch cuts on the $x$-plane $[-2i, -(\sqrt{3}+\ep)i],
[-\sqrt{3}i, \sqrt{3}i]$, and $[(\sqrt{3}+\ep)i , 2i]$ 
before taking the limit $x_1 \to x_0$. 
These three-branch cuts 
are joined at the 
locations of $x= \pm 
\sqrt{3}i$ 
after taking the limit and they reduce to a single-branch cut $[-2i,2i]$. 
Therefore, we can explicitly calculate the values $(2N_0,N_1)$ as 
follows:
$
2N_0= \frac{1}{2\pi i} \oint_{A_0} T(x) dx=
 \frac{2}{2\pi i}\int_{-\sqrt{3}i}^{\sqrt{3}i}\frac{6}{
\sqrt{x^2+4}} dx =4$, and
$N_1=
 \frac{1}{2\pi i} \oint_{A_1} T(x) dx=
\frac{2}{2\pi i}\int_{\sqrt{3}i}^{2i} \frac{6}{
\sqrt{x^2+4}} dx=1
$
as we expected.
}
This choice for  $B_8(x)$ can be written as
a Chebyshev polynomial of degree 3
with appropriate identification.
At these intersection points, the vacua survive 
when the ${\cal N}=2$ theory
is perturbed by a quadratic ($n=0$) superpotential.
By the addition map, the above matrix model curve 
from glueball approach can be obtained also from the pure $USp(4)$
gauge theory ($2N-N_f=4$) with the breaking pattern 
$USp(4) \to USp(2) \times U(1)$ \cite{ao2}.

2) For the second solution $USp(6) \to \widehat{USp(2)} \times U(2)$, 
the matrix model curve 
$\widehat{y_{m}}^2=t \left(t+d \right)^2-
4\eta \La^6$ has an extra double root at 
$d=-3\epsilon \La^2,\ \epsilon^6=1$ and it can be factorized as 
$\widehat{y_{m}}^2=\left(t-\epsilon \La^2 \right)^2
\left(t-4\epsilon \La^2 \right)$ with $\eta=\ep^3$. 
If we simply put the values $(2N_0,N_1,N_f)=(2,2,2)$ 
into our general formula 
(\ref{curve1}) for the 
matrix model curve at the singularities, we obtain the following results,
\begin{eqnarray}
y_{m}^2=x^2\left(x^2+3\eta \La^2 \right)^2+4\eta^3\La^6=
\left(x^2 + \eta \La^2 \right)^2 \left( x^2 +4\eta \La^2 \right), \nonu
\end{eqnarray}
where $\eta$ is $6$-th root of unity. If we identify this $\eta$ with 
$-\epsilon$ in the solution for factorization problem, we see the 
agreement of two approaches.
\footnote{
By using
$
\widehat{y_m}^2=
\left(x^2+4 \right) \left(x^2+ 1\right)^2$, and $
B_8(x)=
x^2 \left(x^6 +6x^4+9x^2+2 \right)$, 
the function $T(x)$ 
is given by
$
T(x)=\frac{6}{
\sqrt{x^2+4}}$.
Therefore, we can explicitly calculate the values $(2N_0,N_1)$ as 
follows:
$
2N_0=
 \frac{2}{2\pi i}\int_{-i}^{i}\frac{6}{
\sqrt{x^2+4}} dx =2$, and
$N_1=
\frac{2}{2\pi i}\int_{i}^{2i} \frac{6}{
\sqrt{x^2+4}} dx=2
$.
}
At the intersection points, the vacua survive 
when the ${\cal N}=2$ theory
is perturbed by a quadratic ($n=0$) superpotential.

$\bullet$ $USp(6)$ with $N_f= 4$

The factorization problem \cite{afo2}
turned out that 
the matrix model curve has an overall $x^2$ factor 
and is classified as a degenerated case
which will be studied later.

$\bullet$ $USp(6)$ with $N_f= 6$

Again the factorization problem \cite{afo2}
turned out 
the matrix model curve has an overall $x^2$ factor 
and it is classified as a degenerated case
which will be studied later.



As in the $SO(N)$ case we can obtain the coupling constant 
by using simple replacement of $(2N_0+2)$ 
in pure gauge theory result with $(2N_0-N_f+2)$ 
we can get the coupling constant 
near the singularity without explicit calculation,
\begin{eqnarray}
\frac{1}{2\pi i}\tau_{ij}=\frac{\partial^2 
{\cal F}_p(S_k)}{\partial S_i \partial S_j}-\delta_{ij}
\frac{1}{N_i}\sum_{l=1}^n N_l \frac{\partial^2 {\cal F}_p(S_k)}
{\partial S_i \partial S_l}-\delta_{ij} \left(\frac{2N_0 -N_f+ 2}{N_i}
\right)
\frac{\partial^2 {\cal F}_p(S_k)}{\partial S_0\partial S_i},
\nonu
\end{eqnarray}
where $ i, j=1,2, \cdots, n$.
In the quartic tree-level superpotential case ($n=1$), 
there is only one coupling constant and it is
given by, 
\begin{eqnarray}
\frac{1}{2\pi i}\tau=
- \frac{i \pi}{16}  \frac{\left(2N_0-N_f+2 \right)^2}{ 
\left(2N-N_f+2 \right)^2} 
\frac{1}{ \log \left(\frac{16}{1-k^{\prime 2}}\right)},
\nonu 
\eea
where $k^{\prime ^2}=
\frac{x_0^2\left(x_2^2-x_1^2\right)}{x_1^2\left(x_2^2-x_0^2\right)}$.
 When the condition $(2N_0-N_f+2)=0$ is satisfied, $\tau$ 
becomes zero. As discussed $SO(N)$ case, 
the asymptotic freedom for $USp(2N_0)$ gauge theory breaks down and 
effectively the situation is the same as $n=0$ case. Our result 
gives a consistency with $n=0$ case.

\section{The curve for degenerated $USp(2N)$ gauge 
theory with massless
flavors}
\setcounter{equation}{0}

\indent

Again, we want to get a matrix model curve for
a degenerated $USp(2N)$ case from the degenerated $SO(N)$ results
discussed in previous section.
From the matrix model curve for $SO(N)$ gauge theory,
$
y_{m,d}^2=\left(x^2+a^2 \right)\left(x^2+b^2\right),
$
by simply replacing 
$(a^2,b^2)$ of $SO(N)$ with $(-a^2,-b^2)$, we obtain the matrix 
model curve for $USp(2N)$ gauge theory. In other words, 
we replace the parameter $m$ of a superpotential in $SO(N)$ with $-m$.
Therefore, our matrix model curve 
\footnote{The relation between 
$W_3^{\prime}(x)$ and $F_4(x)$ can be written as
\begin{eqnarray}
y_{m,d}^2&=& F_4(x) = \left(\frac{W_3^{\prime}(x)}{x} \right)^2+
4F, \ \qquad \qquad \qquad
N_f< 2N+1, \nonu \\
y_{m,d}^2&=& F_4(x) = \left(\frac{W_3^{\prime}(x)}{x} \right)^2+
4F - 4x^2\Lambda^{2}, \ \ \qquad
N_f= 2N+1. 
\nonu
\end{eqnarray}
In particular, when $N_f=2N+1$, the matrix model curve 
$y_{m,d}^2$ has an extra contribution  
from $4x^2 \La^2$. Therefore in this particular case, the curve should be
$y_{m,d}^2= \left(x^2-m- 2\La^2 \right)^2+4F$ and $m+2\La^2=
\frac{K^2-F}{K}$.
\label{parti2}} 
becomes 
\bea
y_{m,d}^2=\left(x^2-m \right)^2+4F,\qquad m=\frac{K^2-F}{K},
\qquad  K\equiv \left[ \frac{(-\La^2)^{2N-N_f+2}}{(-F)^{N_1}} 
\right]^{\frac{1}{2N_0-N_f+2}} 
\label{curve3}
\eea
where we use the notation 
$USp(2N)\to \widehat{USp(2N_0)}\times U(N_1)$ with $2N=2N_0+2N_1$.

As in $SO(N)$ case there is a special case in which the condition 
$(2N_0-N_f+2)=0$ is satisfied and the matrix model curve is given as
\bea
y_{m,d}^2=\left(x^2+D \right)^2-\frac{4\La^4}{\eta^2},\qquad
\eta^{2N_1}=1,\qquad 
\mbox{for} \qquad (2N_0-N_f+2)=0. 
\label{curve4}
\eea

Let us demonstrate these general features (\ref{curve3}) and (\ref{curve4}) 
explicitly by comparing 
them with the results from strong-coupling approach 
obtained in \cite{afo2} already. Let us consider $USp(2N)$ 
gauge theories with $N_f (\le 2N+1)$ flavors where $N=2$ or 3.

$\bullet$ $USp(4)$ with $N_f= 2$

In \cite{afo2}, the first kind solution belongs to the degenerated case. 
To see the equivalence we put the values $(2N_0,N_1,N_f)=(0,2,2)$ 
into our formula
(\ref{curve3}) (Note $2N=2N_0+2N_1$). 
However in this breaking pattern, since we have the relation 
$(2N_0-N_f+2)=0$, the 
matrix model curve (\ref{curve4}) is given by
\begin{eqnarray}
y_{m,d}^2=\left(x^2+D \right)^2-\frac{4\La^4}{\eta^2},\qquad 
\eta^4=1. 
\nonu
\end{eqnarray}
The factorization problem \cite{afo2}
resulted in 
$
\widehat{y_m}^2 = \left(x^2 -\al^2 \right)^2 \mp 4 \La^4   
$. 
The gauge group breaks into $USp(4) \to   \widehat{USp(0)} 
\times U(2)$.
One can see the equivalence between two approaches by identifying 
$\al^2=-D$.

$\bullet$ $USp(4)$ with  $N_f=3$

The first kind of solution in \cite{afo2} is the one corresponding 
to the degenerated case with the breaking pattern 
$\widehat{y_{m}}^2= \left(t+D \right)^2+4F,\;D=\frac{F^2}{\La^6}-
\frac{\La^6}{F}$. The $-\al^2$ corresponds to $D$ here. 
 To see the agreement, we put the values 
$(2N_0,N_1,N_f)=(2,1,3)$ into (\ref{curve3}) 
(the breaking pattern with $(N_0,N_1,N_f)=(0,2,3)$ 
where $USp(4) \to \widehat{USp(0)} 
\times U(2)$ 
reproduces the same result) and get the matrix model curve,
\begin{eqnarray}
y_{m,d}^2=\left(x^2-m \right)^2+4F,\qquad  m=\frac{\La^6}{F}-
\frac{F^2}{\La^6}. 
\nonu 
\end{eqnarray}
Therefore, there is an agreement.

$\bullet$ $USp(4)$ with  $N_f=4$

1) 
The first kind of solution in \cite{afo2} is represented as 
$\widehat{y_{m}}^2=\left(t+d \right)^2+4\eta \La^2 t$, $\eta^2=1$ and 
the breaking pattern is $ \widehat{USp(0)} \times U(2)$. 
The $\al^2$ corresponds to $-d \pm 2 \La^2$. 
If we 
change the parametrization for matrix model curve by $d=
\frac{-F}{\eta \La^2}-\eta \La^2$, it is rewritten as 
$\widehat{y_m}^2= \left(t+\eta\La^2-\frac{F}{\eta\La^2} \right)^2+4F$.
To see the agreement, we put the values 
$(2N_0,N_1,N_f)=(0,2,4)$ into our general 
formula (\ref{curve3}) and get the matrix model curve as
\begin{eqnarray}
y_{m,d}^2=\left(x^2-m \right)^2+4F,\qquad  m=
\frac{\eta F}{\La^2}-\eta \La^2 \nonu 
\end{eqnarray}
where $\eta^2=1$. We see an agreement. 

2) For the other breaking pattern $USp(4) \to
 \widehat{USp(2)} \times U(1)$, we 
check the equivalence by putting the values 
$(2N_0,N_1,N_f)=(2,1,4)$ into (\ref{curve3}). 
In this case 
since the condition $(2N_0-N_f+2)=0$ is satisfied, the matrix model 
curve (\ref{curve4}) 
is given by
\begin{eqnarray}
y_{m,d}^2=\left(x^2+D \right)^2-4\La^4. 
\nonu
\end{eqnarray}
Since the results from the factorization problem is 
$\widehat{y_{m}}^2=\left(t-\al^2 \right)^2-4\La^4$, 
we see a perfect agreement.

$\bullet$ $USp(4)$ with  $N_f=5$

1) 
For the first kind of solution, the matrix model curve can be 
written as $\widehat{y_{m}}^2= \left(t+D \right)^2+4F,D=
\ep \left( \frac{ F^2}{\La^2}\right)^{\frac{1}{3}}-
\ep F\left( \frac{\La^2}{F^2} \right)^{\frac{1}{3}}$ where 
$\ep^3=1$. The $D$ corresponds to $a/2$ in \cite{afo2}. 
We use different parametrization and the 
breaking pattern is $USp(4) \to  \widehat{USp(0)} \times U(2)$.
Putting the values $(2N_0,N_1,N_f)=(0,2,5)$ into (\ref{curve3}) 
with the footnote \ref{parti2} we get the matrix model curve as
\begin{eqnarray}
y_{m,d}^2=\left(x^2-m -2\La^2 \right)^2+4F,\qquad  
m=- \eta
\left( \frac{ F^2}{\La^2}\right)^{\frac{1}{3}}+ \frac{F}{\eta}
\left( \frac{\La^2}{F^2} \right)^{\frac{1}{3}}-2\La^2  \nonu 
\end{eqnarray}
with $\eta^3=1$.
An agreement between two approaches is evident.

2) By putting the values 
$(2N_0,N_1,N_f)=(2,1,5)$ into (\ref{curve3}) with the footnote
\ref{parti2}
where
$USp(4) \to  \widehat{USp(2)} \times U(1)$, we get
\begin{eqnarray}
y_{m,d}^2=\left(x^2-m -2\La^2 \right)^2+4F,
\qquad  
m=
\frac{F}{\La^2}-3 \La^2. \nonu 
\end{eqnarray}
From the results in \cite{afo2}, the second kind of solution can 
be represented as 
$
\widehat{y_{m}}^2= \left(t+D \right)^2+4\La^2 \left(
3\La^2 +m \right) \nonu
$.
If we introduce $F=\La^2 \left( 3\La^2 + m \right)$, 
we get $m= -3\La^2 +\frac{F}{\La^2}$  
and see an agreement.

$\bullet$ $USp(6)$ with  $N_f=2$

The matrix model curve is 
$\widehat{y_{m}}^2= \left(t-\frac{a}{2} \right)^2-
4\epsilon \La^4$, where 
$\epsilon^3=1$ and the breaking pattern is $USp(6) \to
 \widehat{USp(0)} \times U(3)$. 
In this case since the condition $(2N_0-N_f+2)=0$ is satisfied, 
the matrix model curve (\ref{curve4}) 
from glueball approach is given as
\begin{eqnarray}
y_{m,d}^2=\left(x^2+D \right)^2-\frac{4\La^4}{\eta^2},\qquad
\eta^{6}=1. 
\nonu
\end{eqnarray}
Therefore, by identifying $(\frac{-a}{2}, \epsilon)$ with $(D, 
\eta^4)$, we can see the agreement between two results.

$\bullet$ $USp(6)$ with  $N_f=4$

1) 
For the first kind of the solution from the factorization problem 
it was given in \cite{afo2} as $\widehat{y_{m}}^2=
\left(t+D \right)^2+4F,\; 
D=\eta \frac{F^2+\La^8}{\sqrt{F}\La^4}$, where $\eta^4=1$. 
We use  different parametrization to see the equivalence easily. 
By putting the values $(2N_0,N_1,N_f)=(4,1,4)$
where
 $USp(6) \to
 \widehat{USp(4)} \times U(1)$ 
(We also get the same results if we 
put $(2N_0,N_1,N_f)=(0,3,4)$. 
Therefore, we cannot tell which one corresponds 
to $(\frac{1}{2\pi i} \oint_{A_0} T(x) dx,\frac{1}{2\pi i} 
\oint_{A_1} T(x) dx)$ on this branch without doing these computations), 
we obtain the matrix model curve as
\begin{eqnarray}
y_{m,d}^2=\left(x^2-m \right)^2+4F,\qquad
m=\frac{\eta \La^4}
{\sqrt{F}}-\eta^3 \frac{F\sqrt{F}}{\La^4} \nonu
\end{eqnarray}
where $\eta^4=1$. Therefore, the results obtained from two approaches 
are the same.

2) For the second solution, the matrix model curve is 
$\widehat{y_{m}}^2=\left(t+d \right)^2-
4\epsilon \La^4$, where $\epsilon^2=1$ 
and the breaking pattern is given by $USp(6) \to
 \widehat{USp(2)} \times U(2)$. 
In this case since the condition $(2N_0-N_f+2)=0$ is satisfied, the 
matrix model curve (\ref{curve4}) 
from glueball approach is given as
\begin{eqnarray}
y_{m,d}^2=\left(x^2+D \right)^2-\frac{4\La^4}{\eta^2},\qquad 
\eta^{4}=1.\nonu
\end{eqnarray}
Therefore, by identifying $(d, \epsilon)$ with $(D, \eta^2)$, we 
can see the agreement between two results.
This solution is exactly the same 
as the degenerated solution with $USp(4) \to
 \widehat{USp(0)} \times U(2)$ with $N_f=2$: 
We can expect to have this feature from the addition map.

$\bullet$ $USp(6)$ with  $N_f=6$

1) The matrix model curve is  
$\widehat{y_{m}}^2=\left(t+d \right)^2-
4\epsilon \La^2 t$, where $\epsilon^2=1$ 
and the breaking pattern is given by $USp(6) \to
 \widehat{USp(2)} \times U(2)$. 
In this case by putting the values 
$(2N_0,N_1,N_f)=(2,2,6)$ into (\ref{curve3}), 
the matrix model curve from 
glueball approach is given by
\begin{eqnarray}
y_{m,d}^2=\left(x^2-m \right)^2+4F, \qquad 
m=\frac{\eta F}{\La^2}-
\eta \La^2 \nonu
\end{eqnarray}
where $\eta$ is $2$-nd root of unity. Therefore, by identifying 
$F=d\epsilon \La^2-\La^4$ and $\eta=\epsilon$  respectively we can see the 
agreement between two results. These solutions are exactly the same 
as the degenerated solution with $USp(4) \to
 \widehat{USp(0)} \times U(2)$ we have discussed ($N_f=4$): 
We can expect to have this feature from the addition map.

2) For the breaking pattern $USp(6) \to
 \widehat{USp(4)} \times U(1)$,
 the condition 
$(2N_0-N_f+2)=0$ is satisfied  and the matrix model curve is given by
\begin{eqnarray}
y_{m,d}^2=\left(x^2+D \right)^2-4\La^4 \nonu
\end{eqnarray}
which agrees with the one in \cite{afo2}. 
The $D$ here corresponds to $-s_1$ there.
We can see this solution  from the breaking pattern 
$USp(4) \to
 \widehat{USp(2)} \times U(1)$  through the addition map.
\footnote{
For the last breaking pattern $USp(6) \to
\widehat{USp(0)}\times U(3)$, since 
the factorization problem was not solved exactly in \cite{afo2}, we 
cannot check the results explicitly. 
We simply put the results 
from glueball approach,
$
y_{m,d}^2=\left(x^2+m \right)^2+4F$,
and $
m=\frac{\eta 
F^{\frac{3}{4}}}{\La}-\frac{F^{\frac{1}{4}} \La}{\eta },
\eta^8=1$.}

In summary, for the  $USp(2N)$ 
gauge theories with $N_f (\le 2N+1)$ flavors where $N=2,3$,
the solutions 
(\ref{curve3}) and (\ref{curve4}) coincide with the matrix model curve
from the strong-coupling approach except that we could not check the 
$USp(6)$ gauge theory with $N_f=6$ where there exists a breaking pattern 
$USp(6) \to \widehat{USp(0)} \times U(3)$ (in this case, $N_0=0$). 
Since the SW curve
for $USp(2N) \to USp(0) \times U(N)$ has less power of $x^2$
compared with the SW curve
for $USp(2N) \to \widehat{USp(2N_0)} \times U(N_1), N_0 \neq 0$, 
the factorization problem will be more complicated due to the 
existence of many parameters.
Therefore, one expects that our matrix model curve 
(\ref{curve3}) and (\ref{curve4}) will predict 
the solutions for the  degenerated $USp(2N)$ 
gauge theories with $N_f (\le 2N+1)$ massless flavors where $N \ge 4$
and $USp(2N) \to \widehat{USp(2N_0)} \times U(N_1), N_0 \neq 0$.

\section{The curve for degenerated $SO(N)$ gauge theory
with massive
flavors}

\indent

For the massive case, there exists one extra condition 
$W_3^{\prime}(\pm m_f)=0$ 
where $m_f (\neq 0)$ is the mass of flavor which is different from  
the $m$ introduced  
in \cite{ao2} or section 2 as a parameter of a matrix model curve. 
From this condition, the mass of flavor can be written as 
\begin{eqnarray}
m^2_f=-\frac{1}{2}\left(x_0^2+x_1^2+x_2^2 \right),\qquad 
W_3^{\prime}(x)=x\left( x^2-m^2_f\right).
\nonu
\end{eqnarray}
Equivalently, $m_f=\pm i\sqrt{\frac{x_0^2+x_1^2+x_2^2}{2}}$ is pure 
imaginary where the matrix model curve is described by
$y_m^2 =\prod_{i=0}^{2} \left(x^2 +x_i^2 \right)$ 
with $x_0<x_1<x_2$. The fact that 
$m_f$ is pure imaginary implies the flavor-dependent part
${\cal F}_{flavor}$ can be written in terms of the dual period $\Pi_1$ (not
$\Pi_0$) and other term. Therefore, we encounter a complete different story 
when the masses of flavors are nonzero.   
The absolute value $|m_f|$ is greater than $x_0$: $x_0< |m_f|$. 
When we take 
the singular limit $x_1^2\to x_0^2$, the three-branch cuts $[-ix_2,-ix_1],
[-ix_0,ix_0]$ and
$[ix_1,ix_2]$ are 
joined at $x=\pm ix_0$.

As in massless cases, we want to study two topics, 1) the behavior of
$n=0$ and $n=1$ 
singularities and 2) the curve for degenerated case. 
Although for the degenerated 
case, we obtain a general matrix model curve (in the sense that the curve 
holds for  the generic value of $N$), as we will see below, 
there exist some difficulties for 
$n=0$ and $n=1$ singularities which will be
discussed in next section. Thus, in this section 
we concentrate on the degenerated case in which the matrix model 
curve can be written as $y_m^2=x^2\left(x^2+a^2\right)\left(x^2+
b^2\right)$. According to the notation 
above, we simply put the values $(x_0,x_1,x_2)$ as $(0,a,b)$. There are 
two-branch cuts $[-ib,-ia]$ and $[ia,ib]$. The mass of flavors can 
be written as 
$m_f=\pm i\sqrt{\frac{a^2+b^2}{2}}$, as before. Therefore, the position 
of mass is always on the branch cuts: $a<|m_f|<b$. 
Recall that the property 
of the integral \footnote{
One can consider a noncompact cycle $\widetilde{B}_1$ which has 
an intersection number one with $A_1$ compact cycle and meet at
$x=m_f=P$. Then since the closed loop $B_1-\widetilde{B}_1$ does not
contain any pole in the $y_m(x)$, there exists a relation 
$\oint_{B_1-\widetilde{B}_1} y_m dx=0$. Therefore, the dual period 
$\Pi_1 =\frac{1}{2\pi i} \oint_{B_1} y_m d x$ can be written as
$\frac{1}{2\pi i}
\int_{ix_2}^{i \La_0} y_m d x + \frac{1}{2\pi i} 
\int_{i\La_0}^{ix_2} (-y_m) dx$
which will be $\frac{1}{\pi i} \int_{ix_2}^{i\La_0} y_m dx$.
We perform similar computation for the cycle $\widetilde{B}_1$ and obtain
$\frac{1}{\pi i} \int_{P}^{i\La_0} y_m dx$. By changing the upper bound
$i\La_0$ into $\La_0$ with an extra term, we get the relation 
$\int_P^{ \La_0}y_mdx=\int_
{ix_2}^{ \La_0}y_mdx$. },
\begin{eqnarray}
\int_{ix_0}^{ \La_0}y_mdx=\int_P^{ \La_0}y_mdx=\int_
{ix_2}^{ \La_0}y_mdx,\qquad ix_0 < P <ix_2.
\nonu
\end{eqnarray}
In addition, we have to pay attention to the form of 
${\cal F}_{flavor}$. Naively the contributions from ${\cal F}_{flavor}$
is given by
(\ref{Fflavor}). However, to get the phase factor correctly it 
should be as follows and by simple manipulations 
it is written as a dual period $\Pi_1$ plus other term:
\begin{eqnarray}
{\cal F}_{flavor}&=&N_f\int_{ m_f}^{ \La_0}y_{m}dx+N_f\int^{ -m_f}_{- 
\La_0}y_{m}dx=
2\pi i N_f \Pi_1+
2N_f\int_{i \La_0}^{\La_0}y_m dx.
\nonu
\end{eqnarray}
Note that we used the fact that 
$\int_{ ix_2}^{i \La_0}y_{m}dx = \pi i \Pi_1
$. In previous paper \cite{ao2}, 
there was a typo in a dual period $\Pi_1$. 
Therefore, the effective superpotential with massive 
flavors in degenerated branch can be given as
\begin{eqnarray}
W_{\rm eff} & = & 2\pi i \left[ (N_0-2)\Pi_0+
\left(2N_1-N_f \right) \Pi_1 \right]-2\left(N-2-N_f\right)S\log 
\left(\frac{\La}{\La_0} \right) \nonu \\
& & -2N_f\int_{i \La_0}^{\La_0}y_m dx. 
\label{Weff}
\end{eqnarray}

It is ready to calculate the 
equations of motion for the fields $f_0$ and $f_2$, in general.
When we compute the equation of motion for a field  $F$ for degenerated case, 
it is noteworthy 
to observe the following relation under the 
large $\La_0$ limit where $y_m^2 = x^2 \left( x^2 + a^2 \right) 
\left( x^2 + b^2 \right) = x^2 \left[ \left( x^2 -m_f^2 
\right)^2 +4 F \right]$,
\begin{eqnarray}
2N_f\int_{i \La_0}^{\La_0}\frac{\partial y_m}{\partial F}dx \simeq
-4 N_f\log i.
\nonu
\end{eqnarray}
By using previous results given in 
\cite{ao2} \footnote{
There are two relations for the derivatives of dual periods with respect 
to a field $F$: $
2\pi i\frac{\partial \Pi_0}{\partial F}=2 \log \big| 
\frac{4\La^2}{(a+b)^2} \big|$, and 
$ 2\pi i\frac{\partial\Pi_1}
{\partial F}=2 \log \big| \frac{4\La^2}{(b^2-a^2)} \big|. 
$} and this $\log i$-term,
we obtain one equation of motion,
\begin{eqnarray}
(N_0-2) \log \bigg| \frac{4\La^2}{(a+b)^2} \bigg|+ \left(2N_1 -N_f \right)
\log \bigg| 
\frac{4\La^2}{(b^2-a^2)} \bigg| -N_f \log i=0. 
\label{rela}
\end{eqnarray}
Note the last term
for the phase factor which comes 
from the last term in the formula (\ref{Weff}). 
Strictly speaking, the coefficient in the last term (\ref{rela}) 
we got from the glueball approach is 2, not $-1$. 
It is not clear how this arises but 
on the other hand, the strong-coupling approach implies that 
the relation (\ref{rela}) should be correct. 
Therefore, after we solve this equation, 
the matrix model curve 
\footnote{
In particular, when $N_f=N-3$, the matrix model curve 
$y_{m,d}^2$ has an extra contribution  
from $4 \La^2 \left( x^2-m^2 \right)$. 
Therefore, in this particular case, the curve should be
$y_{m,d}^2= \left(x^2-m_f^2- 2\La^2 \right)^2+4F$ and $m_f^2+2\La^2=
-\frac{K^2-F}{K}$.
\label{par}} 
can be represented as
\begin{eqnarray}
y_{m,d}^2=\left(x^2-m^2_f \right)^2+4F,
\qquad \ m^2_f=-\frac{K^2-F}{K},
\qquad K\equiv \left[\frac{(\ep \La^2)^{N-2}( \ep i\La^2)^{-N_f}}{(\eta i 
\sqrt{F})^{2N_1-N_f}} 
\right]^{\frac{1}{N_0-2}}. 
\label{form}
\end{eqnarray}
where $\ep^2=\eta^2=1$.
Recall that $m_f^2 =-\left(a^2 + b^2  \right)/2$ and $4F=
-\left(b^2 -a^2 \right)^2/4$.
The quartic (in $x$) matrix model curve in the breaking pattern 
$SO(N) \to SO(N_0) \times \widehat{U(N_1)}$ with $N_f$ massive flavors
($N=N_0+2N_1$) depends on $N_0,N_1,N_f$ and a field $F$. 
Contrary to the non-degenerated case, this formula is always 
valid (On the non-degenerated case, there exists another condition 
$c<0$ where $c$ is given by (\ref{curvetilde}) or (\ref{curve6}) 
which will be discussed in next section).  Note that we use a 
different notation to see the condition $W_3^{\prime}(\pm m_f)=0$ easily.
When the condition $(N_0-2)=0$ is satisfied, we have to use 
different formula, as we did before,
\begin{eqnarray}
y_{m,d}^2=\left(x^2-m^2_f \right)^2 - \frac{4\La^4}{\eta^2},
\qquad \eta^{2N_1-N_f}=i^{N_f},\qquad \mbox{for} \qquad 
(N_0-2)=0. \label{form1}
\end{eqnarray}
Here we used the equation (\ref{rela}) when $(N_0-2)=0$ and 
identified $\eta$ with
$\pm \frac{4\La^2}{b^2-a^2}$. Note $m_f^2 = -\frac{1}{2} \left( a^2+
b^2 \right)$.

The discussions above are valid to the $r=0$ branch 
studied in \cite{afo1}. 
The immediate question is how do we represent the matrix model curve 
for $r\ne 0$ vacua? 
As already discussed in \cite{csw1}, the nonzero $r$-vacua 
can be realized by changing a mass parameter $m_f$. If the singularity 
passes through the first cut (enclosed by the $A_1$ contour), 
the $A_1$  contour is deformed and
has transformed to 
$A_1+C_1-\widetilde{C}_1$. Note that in our present 
case, the contours 
$A_1$ and $C_1(\widetilde{C}_1)$ correspond to a contour around the branch 
cut $[ix_1,ix_2]$ and a
contour around $m_f$ on the upper(lower) sheet respectively. 
On the pseudo-confining vacua the singularity is located 
on the second sheet, so we have $\frac{1}{2\pi i}
\oint_{\widetilde{C}_1}T(z)dz=0$ and $ 
\frac{1}{2\pi i}\oint_{{C}_1}T(z)dz=1$.
After transition we have $N_1-1=\frac{1}{2\pi i}\oint_{{A}_1}T(z)dz$. 
Thus, the effective value of $N_1$ is reduced by 1 in this 
transition.
Therefore, if we pass through $r$-singularities, 
after transition the effective value of $N_1$ has been reduced by 
$(N_1-r)$. 
In addition, the effective number of flavor 
is also reduced by $(N_f-r)$. The contribution from  
the singularity on the first sheet 
has different sign from the one on the second sheet because 
the sign of $y_m$ is different between the two sheets. 
Therefore, by adding the quantity $r$ from the second sheet,
${\cal F}_{flavor}$
on the $r$-branch can be summed and represented as 
follows:
\begin{eqnarray}
{\cal F}_{flavor}&=& \left(N_f-r \right)\int_{ m_f}^{ \La_0}
y_{m}dx+ \left(N_f-r \right)\int^{ -m_f}_{- 
\La_0}y_{m}dx+r\int_{ m_f}^{ \La_0}\left(-y_{m} \right) dx+
r\int^{ -m_f}_{- 
\La_0} \left(-y_{m} \right) dx \nonu \\
&=&2\widetilde{N_f}\int_{ ix_2}^{ \La_0}y_{m}dx=
2\pi i \widetilde{N_f} \Pi_1 +2\widetilde{N_f}
\int_{ i\La_0}^{ \La_0}y_{m}dx
\nonu
\end{eqnarray}
where we introduce  the following notation,
\begin{eqnarray}
\widetilde{N}=N-2r,\qquad 
\widetilde{N_f}=N_f-2r,\qquad \widetilde{N_1}=N_1-r.
\label{tildevariable}
\end{eqnarray}
Therefore, by combining the modified $N_1$ and the contribution from 
the flavors,
after transition the effective 
superpotential can be given as
\begin{eqnarray}
W_{\rm eff}&=&2\pi i \left[(N_0-2)\Pi_0+ \left(2\widetilde{N_1}-
\widetilde{N_f} \right) \Pi_1 \right]-
2(\widetilde{N}-2-\widetilde{N_f})S\log 
\left(\frac{\La}{\La_0} \right) \nonu \\
& & -
2\widetilde{N_f}\int_{i\La_0}^{\La_0} y_mdx. 
\nonu
\end{eqnarray}
In the log term, we used the fact
$\widetilde{N}-2-\widetilde{N_f}=N-2-N_f$.   
This effective superpotential for nonzero $r$-vacua 
takes the same form for $r=0$ vacua (\ref{Weff}) with 
modified quantities (\ref{tildevariable}) and therefore 
includes (\ref{Weff}). 

Now it is straightforward to obtain  the matrix model curve 
for $r\ne 0$ vacua.  
The general matrix model curve 
\footnote{
When $\widetilde{N_f}=\widetilde{N}-3$, the matrix model curve 
should be
$y_{m,d}^2= \left(x^2-m_f^2- 2\La^2 \right)^2+4F$ and $m_f^2+2\La^2=
-\frac{K^2-F}{K}$.
\label{con}} 
can be given by starting with (\ref{form}) and changing the values
$(N,N_1,N_f)$  appearing in (\ref{form}) 
into $(\widetilde{N},\widetilde{N_1},\widetilde{N_f})$ 
where
they are defined as in (\ref{tildevariable}) : 
\begin{eqnarray}
y_{m,d}^2=\left(x^2-m^2_f \right)^2+4F,
\qquad  
m^2_f=-\frac{K^2-F}{K},
\qquad K\equiv \left[\frac{(\ep \La^2)^{\widetilde{N}-2}( \ep i\La^2)^
{-\widetilde{N_f}}}{(\eta i \sqrt{F})^{2\widetilde{N_1}-\widetilde{N_f}}} 
\right]^{\frac{1}{N_0-2}} 
\label{tildeform}
\end{eqnarray}
where $\ep^2=\eta^2=1$
and moreover when the $N_0$ is equal to 2, 
as we did before, 
the matrix model curve with modified quantities (\ref{tildevariable}) 
becomes
\begin{eqnarray}
y_{m,d}^2=\left(x^2-m^2_f \right)^2-\frac{4\La^4}{\eta^2},
\qquad 
\eta^{2\widetilde{N_1}-\widetilde{N_f}}=i^{\widetilde{N_f}},
\qquad \mbox{for}
\qquad  (N_0-2)=0. 
\label{tildeform1}
\end{eqnarray}
Therefore both equations (\ref{tildeform}) and (\ref{tildeform1}) 
are the most general expressions including 
(\ref{form}) and (\ref{form1}).
To demonstrate these general solutions, let us consider some explicit 
examples, $SO(N)$ theories  
with $N_f (\le N-3)$ flavors for  $N=4,5$, or 6.
For $SO(6)$ gauge theories, 
we only consider even number of flavors as we did in \cite{afo1}.

$\bullet$ $SO(4)$ with $N_f=1$: Non-baryonic $r=0$ branch

1) $SO(4) \to SO(2) \times \widehat{U(1)}$ 

The matrix model curve derived by the factorization problem of
degenerated case was given 
in \cite{afo1}. After the factorization (Note that the $F_6(x)$ 
has an overall factor
$x^2$ in the non-degenerated case) we have obtained $F_4(x)=
\left(x^2-A \right)^2-
4\La^2 \left(x^2-m^2_f \right)$ and  from the relation
between $W_3^{\prime}(x)$ and $F_4(x)$ one can read off the 
$W_3^{\prime}(x)$.  
By the condition $W_3^{\prime}(\pm m_f)=0$, we see $A=m^2_f$ 
and the matrix model curve from strong-coupling approach 
can be written as 
$\widehat{y_m}^2=\left(x^2-m^2_f-2\La^2 \right)^2-4\La^4$.
 
Now let us see this observation from the glueball approach.
If we  substitute 
the values $(N_0,N_1,N_f)=(2,1,1)$ where $N=N_0+2N_1$ 
into our general formula (\ref{form1}) (In this case 
the condition $(N_0-2)=0$ is satisfied and also see the footnote 
\ref{par}), we 
obtain the matrix model curve on the 
degenerated case,
\begin{eqnarray}
y_{m,d}^2=\left(x^2-m^2_f -2\La^2 \right)^2+4\La^4 \nonu
\end{eqnarray}
with the phase factor $\eta=i$.
Therefore, we do not see the agreement:$\widehat{y_m}^2 \neq y_{m,d}^2$. 
Let us go back 
a derivation for (\ref{form1}).
According to the strong-coupling approach, the curve is
characterized by 
$\widehat{y_m}^2=\left(x^2- m_f^2\right)
\left( x^2-m_f^2-4\La^2  \right)$.
Then $a^2$ in the matrix model curve 
$y_{m,d}^2= \left( x^2 + a^2 \right)\left( x^2 + b^2 \right)$ 
corresponds to $-m_f^2-4\La^2$ and $b^2$ corresponds to 
$-m_f^2$. So in this particular case, 
the previous condition $a < |m_f| < b$ does not hold.
In other words, the glueball approach is not allowed. 
\footnote{
The first kind of solution in \cite{afo1} 
where $SO(4) \to \widehat{U(2)}$ 
has the matrix model curve as $\widehat{y_m}^2=\left(x^2-D 
\right)^2+4F$ 
where $D=-b/2$ and $4F=c-b^2/4$ 
in the notation of \cite{afo1}. 
In order to get these results, we 
used two relations: One for $c$ and the other is
$ a^2\left( a+ 9 \La^2\right) + 
3a \left( a + 4\La^2\right)m_f^2 +\left( 3a + 4\La^2  \right)m_f^4
+ m_f^6=0
$ together with $b=-2m_f^2-4\La^2$. 
By using these two relations, one can get an equation for $m_f^2$ or
$D$ in terms of $F$ or $c$. 
To see the equivalence between 
two approaches
(glueball approach and strong-coupling approach), we put $(N_0,N_1,N_f)=
(0,2,1)$ into the general formula (\ref{form}) 
and  get the matrix model curve
$
y_{m,d}^2=\left(x^2-m_f^2- 2\La^2 \right)^2+4F$, and $m_f^2=-\frac{\eta 
F^{\frac{3}{4}}}{\La}+F\frac{\La}{\eta F^{\frac{3}{4}}} + 2\La^2
$ 
where $\eta$ is $4$-th root of unity.
This leads to  a  disagreement. }

$\bullet$ $SO(5)$ with $N_f=1$: Non-baryonic $r=0$ branch

To see the equivalence between two approaches easily, 
we solve the factorization again by using the following parametrization, 
$P^2_4(x)-4x^2 \La^4 \left(x^2-m^2_f \right)=H^2_2(x)\left[(x^2-b)^2+4F
\right]$. 
Note $4F=c$ in the notation of \cite{afo1}. From the 
relationship between $F_4(x)$ and 
$W_3^{\prime}(x)$, we obtain $b=m^2_f$ and additionally there exists 
one equation 
corresponding to (5.15) of \cite{afo1}. However for the 
present purpose, we reexpress it in terms of $F$ by writing $a$ 
as a function of $F$ or $c$,
$F^4-2 F^2 \La^8 + \La^{16}- F \La^8 m_f^4=0$.
Solving this equation, we can get $m^2_f$ as follows:
$
m^2_f=\pm \frac{F^2-\La^8}{\sqrt{F}\La^4}
$. In this case, the matrix model curve is
written as $\widehat{y_m}^2=\left( x^2-m_f^2\right)^2 +4F$. 

On the other hand,
if we put the values $(N_0,N_1,N_f)=(3,1,1)$ where
$SO(5) \to SO(3) \times \widehat{U(1)}$ into (\ref{form}) 
we obtain the matrix model curve,
\begin{eqnarray}
y_{m,d}^2=\left(x^2-m^2_f \right)^2+4F,\qquad m^2_f=\pm 
\frac{\left(\La^8-F^2\right)}{\La^4 \sqrt{F}}. \nonu
\end{eqnarray}
Therefore, there is a perfect agreement.
Although there is a breaking pattern $SO(5) \to SO(1) \times 
\widehat{U(2)}$,
the corresponding matrix model curve does not exist.

$\bullet$ $SO(5)$ with $N_f=2$: Non-baryonic $r=1$ branch

From the solution for the factorization, we have obtained 
$\widehat{y_m}^2=
\left(x^2-s \right)^2-4x^2\La^2$ where we wrote the characteristic
function $P_2(x)$ 
in \cite{afo1} 
as $(x^2-s)$. Taking into account the condition $W_3^{\prime}
(\pm m_f)=0$, we get $m_f^2=s$. If we introduce a new
notation $4F\equiv -4\La^4-4s\La^2$, the mass of flavor can 
be rewritten as $m_f^2=-\La^2-\frac{F}{\La^2}$ and the 
curve becomes $\widehat{y_m}^2=
\left(x^2-m_f^2-2\La^2 \right)^2+4F$. Since the 
breaking pattern of this solution is $SO(5) \to SO(1) \times 
\widehat{U(2)}$, by 
putting 
$(\widetilde{N},\widetilde{N_1},\widetilde{N_f},N_0)=(3,1,0,1)$ 
into our general formula (\ref{tildeform}) (we took $\ep=-1$) 
with the footnote \ref{con}, 
we obtain the matrix model curve,
\begin{eqnarray}
y_{m,d}^2=\left(x^2-m_f^2-2\La^2 \right)^2+4F,\qquad m_f^2=-\La^2-
\frac{F}{\La^2}. \nonu
\end{eqnarray}
Therefore, we see the agreement between two approaches.

$\bullet$ $SO(5)$ with $N_f=2$: Non-baryonic $r=0$ branch

By using the same parametrization as \cite{afo1} with 
$F_4(x) = \left(x^4+4bx^2+2c \right)$ 
we can 
represent $c$ as a function of $a,b$ and $m_f$ and there is  an extra 
equation characterized by,
$ a^2 + m_f^4 + 2 a \left(2 \La^2 +m_f^2 \right)=0
$.
To see the equivalence between two approaches let us  introduce a new 
notation $4F\equiv c-4b^2$ and take into account  the condition 
$W_3^{\prime}(\pm m_f)=0$. Finally, by taking $2b=-m_f^2-2\La^2$ and solving
the two equations (the relation for $c$ and the above equation) 
we get the mass $m_f^2=-3\La^2+
\frac{F}{\La^2}$.
The matrix model curve can be written as 
$\widehat{y_m}^2=\left(x^2-m_f^2-2\La^2 \right)^2+4F$.
If we put $(N_0,N_1,N_f)=(1,2,2)$ where there is 
$SO(5) \to SO(1) \times \widehat{U(2)}$ into 
(\ref{form}) (we took $\ep=-1$) with the footnote \ref{par}, 
we obtain the matrix model curve 
\begin{eqnarray}
y_{m,d}^2=\left(x^2-m_f^2 -2\La^2 \right)^2+4F,\qquad m_f^2=
\frac{F}{\La^2}-3 \La^2. \nonu
\end{eqnarray}
We see an agreement. In this case, for $\left(m_f^2 + 2\La^2 \right)^2 + 
4 F=0$, there is
a Chebyshev branch with $SO(5) \to SO(5)$.

$\bullet$ $SO(6)$ with $N_f=1$: Non-baryonic $r=0$ branch

If we rewrite $c$ in the results of \cite{afo1} as $c=4F$ 
the matrix model curve for degenerated case can be represented as 
$\widehat{y_m}^2=\left(x^2-m^2_f \right)^2+4F$ with one equation,
$
F^6 - 2F^3{\Lambda }^{12} - 4F^2m^4_f{\Lambda }^{12} - 
Fm^8_f{\Lambda }^{12} + {\Lambda }^{24}=0$
which is the last equation of (5.18) in \cite{afo1}. 
Solving this constraint, we can obtain $m^2_f$ as follows:
$
m^2_f=\pm \frac{\left(F^{\frac{3}{2}}+\La^6\right)}
{i\La^{3}F^{\frac{1}{4}}}$,
or $\pm \frac{\left(F^{\frac{3}{2}}-
\La^6\right)}{\La^3 F^{\frac{1}{4}}}$.

To see the agreement with our general formula we simply put 
$(N_0,N_1,N_f)=(4,1,1)$ where there exists $SO(6) \to SO(4) \times
\widehat{U(1)}$ into (\ref{form}) 
we can get the matrix model curve 
\begin{eqnarray}
y_{m,d}^2=\left(x^2-m^2_f \right)^2+
4F,\qquad m^2_f=\pm \frac{\left(\La^6-F^
{\frac{3}{2}}\right)}{F^{\frac{1}{4}}\La^3},\qquad \pm i 
\frac{\left(\La^6+F^
{\frac{3}{2}}\right)}{F^{\frac{1}{4}}\La^3}. \nonu
\end{eqnarray}
This implies a perfect agreement.
For $m_f^4+4F=0$, a Chebyshev branch $SO(6) \to SO(6)$ appears.

$\bullet$ $SO(6)$ with $N_f=2$: Non-baryonic $r=0$ branch

There are two types of solutions in \cite{afo1}, 
one has a $t$ factor 
in $H_2(t)$ which can be written as $tH_1(t)$ 
and the other has no $t$ factor in $H_2(t)$. 

1) $SO(6) \to SO(2)\times \widehat{U(2)}$

First, let us consider the former  case. The matrix model curve given in 
\cite{afo1} is $\widehat{y_m}^2=\left(t-m^2_f \right)^2+4\La^4$ 
by inserting
$a= m_f^2 + 2 \eta \La^2$ into $F_4(x)$. 
By putting $(N_0,N_1,N_f)=(2,2,2)$ into our general formula 
(\ref{form1}) 
we obtain 
the matrix model curve from the viewpoint of glueball approach,
\begin{eqnarray}
y_{m,d}^2=\left(x^2-D \right)^2+4\La^4. 
\nonu
\end{eqnarray}
This leads to an  agreement between two approaches.
When $D^2+4\La^4=0$, there is a Chebyshev branch $SO(6) \to SO(6)$.

2) $SO(6) \to \widehat{U(3)}$ 

Secondly, we move to the latter case. The solutions were displayed 
in equation (5.22) of \cite{afo1}. If we use a new 
notation $F\equiv - \left(b+\La^2 \right)^2$, 
the matrix model curve can be rewritten 
as $\widehat{y_m}^2=\left(x^2-m^2_f \right)^2+4F$ with $m^2_f=\pm i
 \frac{\left(F+\La^4\right)}{\La^2}$. 
To see the equivalence with our general 
formula, we put the breaking pattern on this solution, namely
$SO(6) \to \widehat{U(3)}$ 
into the formula (\ref{tildeform}) and get 
the matrix model curve 
\begin{eqnarray}
y_{m,d}^2=\left(x^2-m^2_f \right)^2+4F,\qquad 
m^2_f=\pm i \frac{\left(F+\La^4\right)}{\La^2}. 
\nonu
\end{eqnarray}
One can see an agreement between two approaches.

$\bullet$ $SO(6)$ with $N_f=2$: Non-baryonic $r=1$ branch

1) $SO(6) \to \widehat{U(3)}$ 

The first kind of solution studied in \cite{afo1} gives a relation for
 mass 
$m_f^2=a-2\eta \La^2$. There was a  typo in the equation of \cite{afo1}.
 If we introduce a new notation $4F \equiv 4a\eta \La^2-4\La^4$, the 
mass can be rewritten as 
$m_f^2=\eta \left(\frac{F}{\La^2}-\La^2 \right)$. 
Since this solution has the breaking pattern $SO(6) \to 
\widehat{U(3)}$ by putting 
$(\widetilde{N},\widetilde{N_1},\widetilde{N_f},N_0)=
(4,2,0,0)$ into our general formula (\ref{tildeform}) 
we obtain 
the matrix model curve 
\begin{eqnarray}
y_{m,d}^2=\left(x^2-m_f^2 \right)^2+4F ,\qquad
m_f^2=-\epsilon 
\left(\frac{F}{\La^2}-\La^2 \right) \nonu
\end{eqnarray}
where $\epsilon$ is $2$-th root of unity. This leads to an agreement.

2) $SO(6) \to SO(2)\times \widehat{U(2)}$ 

The other solution has the matrix model curve $\widehat{y_m}^2=
\left(x^2-m_f^2 \right)^2-4\La^4$. Since the condition
$(N_0-2)=0$ is satisfied, the matrix model curve 
from (\ref{tildeform1})
can be rewritten as 
\begin{eqnarray}
y_{m,d}^2=\left(x^2-m_f^2 \right)^2 - 4\frac{\La^4}{\eta^2}, 
\qquad \eta^2=1. 
\nonu
\end{eqnarray}
A perfect agreement occurs.

$\bullet$ $SO(6)$ with $N_f=3$: Non-baryonic $r=0$ branch

The solution for the breaking pattern $SO(2)\times \widehat{U(2)}$ 
can be represented as follows:
$\widehat{y_{m}}^2=\left(x^2-m_f^2-2\La^2 \right)^2-
4\La^4$. We 
use 
$F=\La^2\left(D+m_f^2+\La^2 \right)$. From the relationship 
between $F_4(x)$ and $W_3^{\prime}(x)$, 
we have $D=-m_f^2-2\La^2$ implying 
$F=-\La^4$. 
Putting 
this relation we get the above matrix model curve. 

On the 
other hand, by substituting 
the values $(N_0,N_1,N_f)=(2,2,3)$ where $SO(6) \to 
SO(2) \times \widehat{U(2)}$ 
into (\ref{form1}) 
we obtain the matrix model curve as
\begin{eqnarray}
y_{m,d}^2=\left(x^2+D \right)^2+4\La^4. 
\nonu
\end{eqnarray}
However, as we have discussed in $SO(4)$ with $N_f=1$ case,
the glueball approach given by (\ref{form1})
does not include this case also: The location of the flavors
is different from the region belonging to the branch cut.
For $D^2-4\La^4=0$, $SO(6) \to SO(6)$ Chebyshev branch occurs.

$\bullet$ $SO(6)$ with $N_f=3$: Non-baryonic $r=1$ branch

In \cite{afo1}, it was discussed by using the addition map that this 
branch was  the same as the one for $SO(4)$ with $N_f=1$ 
case. Since our general formula also has the same structure, the glueball 
approach here reproduces the same results from the factorization problem
there.
That is, for the breaking pattern, $SO(6) \to \widehat{U(3)}$, the values
$(\widetilde{N_0}, \widetilde{N_1}, \widetilde{N_f})=(0,2,1)$ correspond to
$(N_0,N_1,N_f)=(0,2,1)$ for  $SO(4)$ theory with $N_f=1$. 
Moreover, on the breaking pattern $SO(6) \to SO(2) \times \widehat{U(2)}$,
 the values
$(\widetilde{N_0}, \widetilde{N_1}, \widetilde{N_f})=(2,1,1)$ 
correspond to $(N_0,N_1,N_f)=(2,1,1)$ of 
  $SO(4)$ theory with $N_f=1$. 

In summary, for the  $SO(N)$ 
gauge theories with $N_f (\le N-3)$ flavors where $N=4,5$ or $6$,
we have checked that the solutions 
(\ref{form}), (\ref{form1}), (\ref{tildeform}), and (\ref{tildeform1}) 
coincide with the matrix model curve
from the strong-coupling approach except that we could not check the 
$SO(4)$ gauge theory with $N_f=1$ where there exists a breaking pattern 
$SO(4) \to \widehat{U(2)}$ (in this case, $N_0=0$) and
$SO(6)$ with $N_f=3$ where there exists a breaking pattern 
$SO(6) \to \widehat{U(3)}$ ($N_0=0$). 
Therefore, at least one expects that our matrix model curve 
(\ref{form}), (\ref{form1}), (\ref{tildeform}), and (\ref{tildeform1}) 
will predict 
the solutions for the  degenerated $SO(N)$ 
gauge theories with $N_f (\le N-3)$ massive flavors where $N \ge 7$
and $SO(N) \to SO(N_0) \times \widehat{U(N_1)}, N_0 \neq 0$.
\footnote{
Using this solution from the glueball approach, one obtains
the matrix model curve for arbitrary $N$. For example, 
for $SO(7)$ with $N_f=1$ (non-baryonic $r=0$ branch),
when we consider
the breaking pattern $SO(3)\times \widehat{U(2)}$, by putting 
$(N_0,N_1,N_f)=(3,2,1)$ into our general formula (\ref{form}) 
we predict the 
matrix model curve as,
$y_{m,d}^2=\left(x^2-m_f^2 \right)^2+4F$, and $  m_f^2=\pm 
\frac{\left(\La^{16}-F^4\right)}{F\sqrt{F}\La^8} 
$.
For $SO(7)$ with $N_f=2$ (non-baryonic $r=1$ branch),
after we put
$(\widetilde{N},\widetilde{N_1},\widetilde{N_f},N_0)=(5,1,0,3)$ into 
(\ref{tildeform})
we 
obtain the matrix model curve,
$y_{m,d}^2=\left(x^2-m_f^2 \right)^2+4F$, and 
$m_f^2=-\frac{\La^6}{F}+\frac{F^2}{\La^6}$ (In this case, we took 
$\ep=-1$).
Moreover, for $SO(7)$ with $N_f=3$ (non-baryonic $r=1$ branch),
the factorization problem becomes the same as the 
one for $SO(5)$ with $N_f=1$ case through the addition map. } 
%
%
%

\section{Discussion on 
the $n=0$ and $n=1$ singularity  
with massive flavors}


\indent

To obtain a general matrix model curve for $n=0$ and 
$n=1$ singularities for our gauge theories, 
we have to describe the behavior of the matrix model curve 
$y_m^2=\prod_{i=0}^{2} \left(x^2+x^2_i \right)$ in 
the limit $x_1 \to x_0$. Contrary to the degenerated case we have studied 
in previous section, there exist two 
possibilities for the position of the mass of flavors. 
One is $x_0 < |m_f| <x_2$ and the other is 
$x_2 < |m_f| <\sqrt{\frac{3}{2}}x_2$. In the former case, 
since the position of flavor mass is on the branch cut $[ix_1,ix_2]$,
${\cal F}_{flavor}$ can be written in terms of a dual 
period $\Pi_1$. Then by following the derivation in previous 
section 
\footnote{
Note that when we compute the equation of motion for $f_0$,
since the extra piece from the flavor part, an 
integral $ \int_{i \La_0}^{\La_0}
\frac{\partial y_m}{\partial f_0}dx$,  
behaves like ${\cal O}(\La_0^{-2})$,
it does not contribute. 
Then the equation of motion for $f_0$ looks similar to the pure case
\cite{ao2}
and it is given by
$
\frac{x_2^2-2x_0^2}{x_2^2} = -\cos \left[\frac{\pi 
(2 \widetilde{N_1}-\widetilde{N_f})}{\widetilde{N}-
\widetilde{N_f}-2} \right]
$
and the equation of motion for $f_2$ leads to
$
\left(\widetilde{N}-\widetilde{N_f}-2 \right) 
\log \big|\frac{x_2^2}{4\La^2} \big| + N_f 
\log i =0.
$
After solving this, one gets
$
x_2^2 = \pm 4 \eta \La^2$, and $ \eta^{\widetilde{N}-\widetilde{N_f}-2} 
=i^{\widetilde{N_f}}.
$}
 we obtain the matrix model curve near the singularity on the 
non-degenerated case as follows:
\begin{eqnarray}
y_m^2&=&x^2 \left(x^2-m_f^2 \right)^2+
\langle f_2 \rangle x^2 +\langle f_0 \rangle =
\left[ x^2 \pm 2 \eta \La^2 \left(1+c \right) \right]^2 \left( x^2 
\pm 4 \eta \La^2 \right), \nonu
\end{eqnarray}
where the fields, parameter of a superpotential, 
and glueball field are given by 
\begin{eqnarray}
\langle f_2 \rangle &=& 4\eta^2 \La^4 \left(1+2 c 
\right),\qquad 
\langle f_0 \rangle=\pm 16\eta^3 \La^6 \left(1+c 
\right)^2, \nonu \\
m_f^2 &=& \pm 2\eta \La^2 \left(2+ c \right), \qquad  
\langle S \rangle=-\eta^2\La^4 \left(1+2 c 
\right),  \label{curvetilde}  \\
c&\equiv& \cos \left[ \frac{\pi (2\widetilde{N_1}-\widetilde{N_f})}
{\widetilde{N}-\widetilde{N_f}-2} \right],
\qquad \eta^{\widetilde{N}-
\widetilde{N_f}-2}=i^{\widetilde{N_f}} \nonu
\end{eqnarray}
together with (\ref{tildevariable}).
We can see that the solutions
for the roots $x_2^2=\pm 4\eta \La^2$
and $x_0^2=\pm 2\eta \La^2 \left(1+c \right)$
satisfy the relation $2x_0^2\le x_2^2$, implying that these solutions are 
consistent with the condition we have assumed before (Note $|m_f|=
\sqrt{(2x_0^2+x_2^2)/2} < x_2$).
Note that $c$ should be negative or zero 
but $c\neq -1$. When $c=-1$, then this 
will be the vacuum in the degenerated case. 
The constraint $|m_f| < x_2$ is too restrictive and most of the 
examples from factorization problem \cite{afo1,afo2} 
are beyond this criterion.   

There exists one more possibility, $x_2 < |m_f| <\sqrt{\frac{3}{2}}x_2$. 
In this case, since ${\cal F}_{flavor}$ cannot be written in 
terms of a dual period $\Pi_1$, we cannot obtain a general matrix 
model curve. Namely, the integral ${\cal F}_{flavor}$ 
can be represented by an elementary function but when we 
solve the equations to obtain the roots $x_0$ and $x_2$, 
a general solution for 
the  matrix model curve is not apparent. 

In summary, at $n=0$ and $n=1$ singularity we obtain some partial results 
for matrix model curve (the curve does not exist for all the range of 
$N, N_1$, and $N_f$ but for some particular values) in which 
the condition, $c$ is negative 
or zero ($c\neq -1$), 
should be 
satisfied. To see the correctness of the result let us  study an example 
satisfying this condition only. 
\footnote{
For $SO(5)$ with $N_f=2$ (non-baryonic $r=0$ branch),
the matrix model curve  was given 
in \cite{afo1} as $\widehat{y_m}^2=\left(x^2-m^2_f \right)^2
\left(x^2-4 \La^2 \right)$.  
To see whether there exists the equivalence of two approaches,
we substitute 
$(N_0,N_1,N_f)=(3,1,2)$ into our general formula (\ref{curvetilde}) 
and then 
obtain the matrix model curve: 
$
y_m^2=\left( x^2 \pm 2 \eta \La^2 \right)^2
\left(x^2 \pm 4\eta \La^2 \right)
$
where $\eta=-1$ with $m_f^2=\pm 4 \eta \La^2$.
However, in this case, $c=1$. Therefore, we cannot
see an agreement between two approaches. 
That is, the matrix model curve exists only from the strong-coupling 
approach. 
One can easily 
see that for $SO(6)$ with $N_f=1$ (non-baryonic $r=0$ branch),
the $c$ value is $1/2$, for $SO(6)$ with $N_f=2$ 
(non-baryonic $r=0$ branch), $c=1$, and for 
$SO(6)$ with $N_f=3$ (non-baryonic $r=0$ branch) 
the $c$ value is $-1$. Most of the examples  studied in \cite{afo1}
belong to the case where $c$ is positive or $-1$.}

$\bullet$ $SO(5)$ with $N_f=1$: Non-baryonic $r=0$ branch

The matrix model curve derived by the factorization was given 
in \cite{afo1} as (There is a typo in the presentation of 
\cite{afo1}) $\widehat{y_m}^2=x^2\left(x^2-m^2_f \right)^2-
4\La^4\left(x^2-m^2_f\right)$. 
This matrix model curve has an extra double root at 
$m^2_f=4\eta i \La^2$ where $\eta^2=1$ and have the 
following factorized form, 
$\widehat{y_m}^2=\left(x^2 \pm  2i\La^2 \right)^2\left(x^2 \pm
4i \La^2\right)$. 
To see the equivalence of two approaches we substitute 
$(N_0,N_1,N_f)=(3,1,1)$ where $SO(5) \to SO(3) \times 
\widehat{U(1)}$ and $N=N_0+2N_1$ 
into our general formula (\ref{curvetilde})   
and then 
obtain the matrix model curve: 
\bea
y_m^2=\left( x^2\pm2 \eta \La^2 \right)^2
\left(x^2\pm 4\eta \La^2 \right)
\nonu
\eea
where $\eta^3=i$. Therefore for pure imaginary $\eta=-i$, 
we see the agreement.



For $USp(2N)$ case, as already discussed in section \ref{USpmassless}, 
we have to pay 
attention to the signs in the matrix model curve. To get the right 
results for 
$USp(2N)$ theory  from the results for $SO(N)$ theory, we 
simply replace $x_0^2 \to -x_0^2, x_2^2 \to -x_2^2$, and $(N-2) \to (2N+2)$ 
respectively.
With this in mind, the matrix model can be written as
\begin{eqnarray}
y_m^2&=&x^2 \left(x^2-m_f^2 \right)^2+
\langle f_2 \rangle x^2 +\langle f_0 \rangle =
\left[ x^2 \pm 2 \eta \La^2 \left(1+c \right) \right]^2 \left( x^2 
\pm 4 \eta \La^2 \right), 
\nonu
\end{eqnarray}
where the fields, parameter of a superpotential, 
and glueball field are given by 
\begin{eqnarray}
\langle f_2 \rangle &=& 4\eta^2 \La^4 \left(1+2 c 
\right),\qquad 
\langle f_0 \rangle=\pm 16\eta^3 \La^6 \left(1+c 
\right)^2, \nonu \\
m_f^2 &=& \pm 2\eta \La^2 \left(2+ c \right), \qquad  
\langle S \rangle=-\eta^2\La^4 \left(1+2 c 
\right),\label{curve6} \\
c&\equiv& \cos \left[ \frac{\pi (2\widetilde{N_1}-\widetilde{N_f})}
{2\widetilde{N}-\widetilde{N_f}+2} \right],
\qquad \eta^{2\widetilde{N}-
\widetilde{N_f}+2}=i^{\widetilde{N_f}}, \nonu
\eea
with the modified quantities
are given by $2 \widetilde{N} = 2N -2r, \widetilde{N_f}=N_f-2r$, and $
2\widetilde{N_1}=2N_1-2r$.
In addition, we can see that the solutions for different
parametrization are given by  
$
x_2^2=\pm 4\eta \La^2$,
and $x_0^2=\pm 2\eta \La^2 \left(1+c \right)$
which satisfy the relation $2x_0^2\le x_2^2$, 
implying that these solutions are 
consistent with the condition we have assumed before.
Note that $c$ defined in (\ref{curve6}) 
should be negative or zero ($c\neq -1$). 
\footnote{
For $USp(4)$ with $N_f=2$ (non-baryonic $r=1$ branch),
the matrix model 
curve is $\widehat{y_m}^2=
t^3+2\left(m_f^2-s_1\right)t^2+
\left(-4\La^4+m_f^4-2m_f^2s_1+s_1^2\right)t-4\left(\La^4m_f^2-
\La^4s_1\right)$ 
with $a=-2m_f^2$. Requiring an extra double root, the matrix model
curve  becomes 
$\widehat{y_m}^2=\left(t \pm 2i\La^2 \right)^2
\left(t\pm 4i\La^2\right)$ at $m_f^2=\mp 4i\La^2$. 
To see whether the equivalence arises,  let us put $(
\widetilde{N},\widetilde{N_1},\widetilde{N_f})=(1,0,0)$ 
into (\ref{curve6}) where the breaking pattern is 
$ USp(4) \to \widehat{U(1)} \times USp(2)
$
we obtain the 
matrix model as
$
y_m^2=\left(x^2\pm 4\eta 
\La^2 \right)^3, \eta^{4}=+1$.
In this case, since $c=1$, one cannot see the agreement between 
two approaches.
In other words, there exists only a matrix model curve from 
strong-coupling approach.
One can easily 
see that for $USp(6)$ with $N_f=2$ (non-baryonic $r=0$ branch),
the $c$ value is $1/2$, for $USp(6)$ with $N_f=4$ 
(non-baryonic $r=0$ branch), $c=1$, and for 
$USp(6)$ with $N_f=6$ (non-baryonic $r=0$ branch) 
the $c$ value is $-1$. Most of the examples  studied in \cite{afo2}
belong to the case where $c$ is positive or $-1$. }
Let us consider one example.

$\bullet$ $USp(4)$ with $N_f=2$: Non-baryonic $r=1$ branch

The matrix model 
curve  can be written as $\widehat{y_m}^2=t^3-2m_f^2 
t^2+ \left(-4\La^4 + m_f^4 \right) t+ 4 \La^4 m_f^2 $ 
with $a=-2m_f^2$. Requiring an extra 
double root it becomes $\widehat{y_m}^2=\left(t 
\pm 2i\La^2\right)^2\left(t\pm 
4i\La^2\right)$ at $m_f^2=\mp 4i\La^2$. To see the equivalence  we put 
the values $(2\widetilde{N},2\widetilde{N_1},
\widetilde{N_f})=(2,2,0)$ where 
$USp(4) \to USp(0) \times \widehat{U(2)}$
into (\ref{curve6}) 
we obtain the matrix model 
\begin{eqnarray}
y_m^2=\left(x^2 \pm 2 \eta \La^2 \right)^2
\left(x^2\pm 4\eta \La^2 \right), 
\qquad \eta^{4}=1. \nonu
\end{eqnarray}
For the cases of pure imaginary $\eta=\pm i$, the matrix model curve 
from the glueball approach coincides with the curve from strong-coupling 
approach.

%

\vspace{1cm}
\centerline{\bf Acknowledgments}

This research of CA was supported by Korea Research 
Grant(KRF-2002-015-CS0006). This research of YO was supported by a 21st 
Century COE Program at TokyoTech "Nanometer-Scale Quantum Physics" by the
Ministry of Education, Culture, Sports, Science and Technology. 
CA is grateful to B. Feng for discussions.
YO wants to thank for Hiroyuki Fuji, Katsushi Ito, Shun'ya Mizoguchi and 
Masaki Shigemori for  discussions, and Yukawa Institute for Theoretical Physics where the final stage of this work was undertaken.

\end{document}